\documentclass[reprint,pre,amsmath,amssymb,aps,showkeys,superscriptaddress]{revtex4-1}

\usepackage{graphicx}
\usepackage{dcolumn}
\usepackage{bm}
\usepackage{color,soul}
\usepackage{breqn}
\usepackage{epstopdf}
\usepackage{img_macros}
\usepackage{overpic}

\graphicspath{{./}{figs/}{traj_snaps/}}

\makeatletter
\let\cat@comma@active\@empty
\makeatother

\begin{document}

\preprint{APS/123-QED}

\title{\Large Nematohydrodynamics for Colloidal Self-Assembly and Transport Phenomena}

\author{Sourav~Mondal}
\affiliation{Mathematical Institute, University of Oxford, Oxford OX2 6GG, UK}

\author{Apala~Majumdar}
\affiliation{Department of Mathematical Sciences, University of Bath, Bath BA2 7AY, UK}

\author{Ian~M.~Griffiths}
\affiliation{Mathematical Institute, University of Oxford, Oxford OX2 6GG, UK}
\email{ian.griffiths@maths.ox.ac.uk}

\date{\today}

\begin{abstract}

We study colloidal particles in a nematic-liquid-crystal-filled microfluidic channel and show how elastic interactions between the particle and the channel wall lead to different particle dynamics compared with conventional microfluidics. For a static particle, in the absence of a flow field, the director orientation on the particle surface undergoes a rapid transition from uniform to homeotropic anchoring as a function of the anchoring strength and the particle size. In the presence of a flow field, in addition to fluid viscous stresses on the particle, nematic-induced elastic stresses are exerted as a result of the anchoring conditions. The resulting forces are shown to offer the possibility of size-based separation of individual particles. For multi-particle systems, the nematic forces induce inter-particle attraction induces particle attraction, following which the particles aggregate and reorientate. These results illustrate how coupled nematic-hydrodynamic effects can affect the mobility and spatial reorganization of colloidal particles in microfluidic applications.

\end{abstract}

\keywords{nematic fluid; microchannel; Beris--Edwards; particle dynamics}

\maketitle

\section{\label{sec:level1} Introduction}

Nematic liquid crystals (NLCs) are important examples of  complex anisotropic fluids with special distinguished directions \cite{lagerwall2012new}. NLCs combine the intrinsic fluidity of liquids with long-range orientational ordering of the constituent rod-like molecules. The orientational order couples with the flow and induces novel effects compared with isotropic Newtonian fluids, such as backflow, anisotropic stresses and multiple viscosities.
The study of NLCs in microfluidic environments is relatively new, with substantial experimental interest  since around 2011. Subsequently, experimentalists have highlighted the immense potential of NLC microfluidics for transport, mixing and particle separation \cite{Sengupta_PRL, Lavrentovich}, while the ability of NLCs to spontaneously organize micron-size particles into regular patterns shows great promise~\cite{lin2008ordering}.  For example, it is possible to generate defect or disclination lines in a NLC microfluidic set-up with an appropriate choice of boundary conditions, material parameters, temperature and flow effects and these defect lines can naturally attract colloidal particles or micro-cargo, which are subsequently transported along these lines as self-assembled chains \cite{Lavrentovich}. Further, the forces facilitating spatial-reorganization of colloidal dispersions in a NLC medium are two to three orders of magnitude higher than in water-based colloids~\cite{vskarabot2008interactions, muvsevivc2006two}. 

Our work is largely motivated by the experiments conducted by Sengupta et~al.~in \cite{Sengupta_PRL}. Here the authors study a NLC microfluidic set-up experimentally and numerically in three different flow regimes: weak, medium and strong, and report on both the flow profiles and the averaged local molecular alignment profiles, referred to as ``director" profiles in the continuum modelling literature. The top and bottom surfaces are treated to induce  homeotropic boundary conditions, so that the nematic molecules are preferentially aligned along the normal to the boundary surfaces, or equivalently the continuum ``director" is parallel to the normal to the boundary. A flow is induced by applying a pressure gradient at the inlet and the observations seem to be invariant across the width of the cell, rendering it to be a largely two-dimensional problem. The weak regime in \cite{Sengupta_PRL} is dominated by the elastic effects of the long-range orientational order with a non-Poiseuille flow profile. The medium regime shows complex coupling between the two effects and the flow profile has an interesting double-peak structure across the height of the cell. The usual Poiseuille flow is observed in the strong flow regime. The director profile switches from a bend-type V state to a splay-type H profile as the flow strength increases and the effect of variations in temperature are also investigated.

We model the motion of particles in the experimental framework reported in Sengupta et al. \cite{Sengupta_PRL}, using the hydrodynamic equations for nematodynamics reported in \cite{giomi2012banding}. We model a cross-section of the channel, which includes the top and bottom channel surfaces and the flow direction, as a rectangle with normal boundary conditions for the nematic director at the top and bottom. The nematic molecules are free at the lateral sides and the flow is induced by a pressure gradient across the length of the rectangle with no-slip conditions on the walls, consistent with experiments. We introduce spherical particles into the channel and the nematic director is preferentially anchored normal to the particle boundaries, although we allow for variable anchoring strength on the particle boundaries. Our manuscript focuses on three separate aspects: (i) a static particle at the centre with variable anchoring strength on its boundary, (ii) the forces experienced by a particle inclusion due to  hydrodynamic effects, nematic stresses and attractive forces induced by the boundary conditions and (iii) the dynamics of two and three particles in a NLC microfluidic environment including the transient dynamics.

We mathematically model this NLC microfluidic environment using the nematodynamics formulation as used in \cite{giomi2012banding}. The state of nematic alignment is described by a two-dimensional (2D) Landau-de Gennes (LdG) $\mathbf{Q}$-tensor, which is a symmetric traceless two-by-two matrix with two degrees of freedom: an angle $\theta$ that describes the preferred in-plane alignment of the nematic molecules or the direction of the nematic director $\ve{n}$, and a scalar order parameter, $s$, that is a measure of the degree of alignment about the director $\ve{n}$. For example, vanishing $s$ corresponds to no local molecular ordering and is often a signature of optical defects. 
We note that the general LdG approach uses a three-dimensional LdG $\mathbf{Q}$-tensor with five degrees of freedom, but our assumption is compatible with in-plane alignment of the nematic molecules i.e., assuming that the nematic molecules are confined to the plane of our 2D domain and a 2D flow profile. A 2D LdG approach has been successfully used for modelling certain planar bistable liquid crystal devices, and the interested reader is referred to \cite{tsakonas2007multistable,kusumaatmaja2015free, luo2012multistability, robinson2017molecular} for some examples. 

We assume an incompressible flow field, consistent with Sengupta et al.'s~\cite{Sengupta_PRL} experiments and the nematodynamic model is defined by evolution equations for both the flow field and the LdG $\mathbf{Q}$-tensor. There are nonlinear coupling terms between the flow field and the $\mathbf{Q}$-tensor in both evolution equations, with new nematic stresses not encountered in conventional fluids.

We numerically investigate how the spherical inclusions interact with the NLC environment without and with flow, for both static and moving particles. The first example concerns a static particle in the NLC microfluidic cell with no fluid flow. For a given anchoring strength on the particle boundary, we study the director profile around the spherical inclusion as a function of the particle size and reciprocally, for a given particle size, we investigate the surrounding director profile as a function of anchoring strength. In both cases, there is a narrow window of parameters within which the director orientation on the particle boundary switches from uniform to normal/homeotropic and we numerically explore the switch in different cases. We then systematically study the force experienced by the particle including the effects of a flow field, the anchoring coefficient $W$, and the particle size. In particular, for a given $W$ and flow field parameter, there is a critical particle size such that, in contrast to conventional liquids, the force attains a maximum at this critical particle size and actually decreases for larger particles owing to the attractive forces exerted by the boundaries, We conclude by numerically studying the motion of two and three colloidal particles in the microfluidic channel, including the transient re-alignment dynamics, how the particles get attracted to each other and are transported through the channel as an agglomerate. 

The paper is organized as follows. In Section II, we review the theory, the relevant non-dimensionalizations, the force exerted on colloidal inclusions in a NLC microfluidic set-up and the numerical methods. We present the results in Section III and the conclusions in Section IV.

\section{\label{sec:level2} Theory}

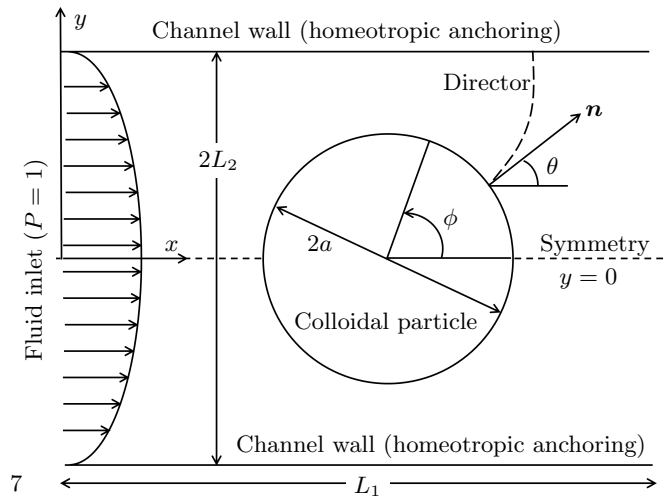
\begin{figure}
\centering
7\begin{overpic}[width = 0.48\textwidth]{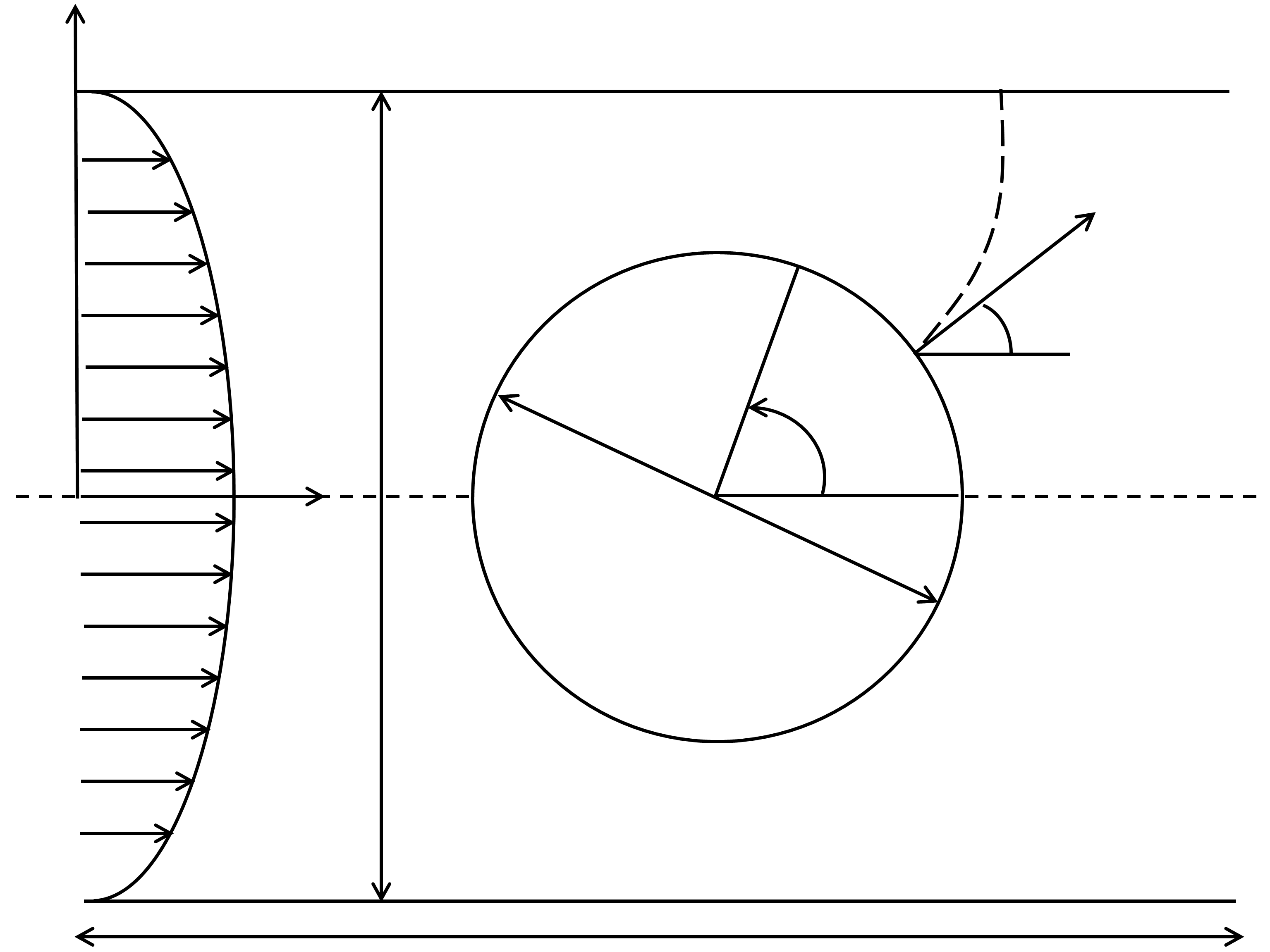}
\put(8,72){$y$}
\put(22,37){$x$}
\put(42,25){Colloidal particle}
\put(44,37){$2a$}
\put(65,41){$\phi$}
\put(87,58){$\ve{n}$}
\put(26,50){\colorbox{white}{$2 L_2$}}
\put(50,-0.4){\colorbox{white}{$L_1$}}
\put(20,70){Channel wall (homeotropic anchoring)}
\put(33,6){Channel wall (homeotropic anchoring)}
\put(81,49){$\theta$}
\put(80,37){Symmetry}
\put(83,32){$y = 0$}
\put(-1,20){\colorbox{white}{\rotatebox{90}{Fluid inlet ($P = 1$)}}}
\put(65,62){Director}
\end{overpic}
\caption[.]{Schematic of the problem definition and system geometry.}
\label{fig.problem.schematic}
\end{figure}

We consider a 2D microfluidic channel (parallel-plate geometry) filled with NLC as shown in Fig.~\ref{fig.problem.schematic}. The director orientation is labelled by the angle $\theta$ so that the nematic director $\ve{n} = \left(\cos\theta, \sin\theta \right)$, and the spherical particle boundary is parameterized by the angle $\phi$, where both angles are defined with respect to the horizontal axis. We apply strong homeotropic anchoring conditions on the channel walls (modelled by Dirichlet conditions) while the anchoring conditions on the spherical colloidal particle are varied from weak to strong in terms of an anchoring coefficient. The fluid flow in the device is driven by an external pressure difference and is also influenced by the stress due to the NLC orientational ordering. We impose no slip and no penetration boundary conditions on the channel walls and particle surface. 

We first numerically compute the director profile inside the channel in the absence of a flow field and systematically study how this varies with anchoring strength $W$, on the particle surface. 
A substantial amount is known about static spherical inclusions inside NLC environments and we provide new quantitative insight into how the director orientation effectively switches from uniform (no anchoring) to normal (strong anchoring) within a small range of values of $W$ and particle sizes. We then consider the force experienced by the spherical particle (held in place at the channel centre) due to hydrodynamic drag and the NLC stress acting on the particle surface, again as a function of flow, size and $W$; and dynamics of a single as well as multiple particles when these are allowed to migrate from the predefined initial location. 

The flow hydrodynamics are described by the incompressible Navier--Stokes equations with an additional stress ($\ve{\sigma}$) due to orientational ordering of the NLC~\cite{giomi2012banding,chen2017global},
\begin{align} 
\label{eq.continuity}
	\nabla \cdot \ve{u} &= 0, \\
\label{eq.flow.vec}
	\rho \lr{\pdf{\ve{u}}{t} + \ve{u} \cdot \nabla \ve{u}} &= -\nabla p + \nabla \cdot ( \mu [\nabla\ve{u} + (\nabla\ve{u})^\prime] + \boldsymbol{\sigma}).
\end{align}
Here $\nabla = \Big( \frac{\partial}{\partial x}, \frac{\partial}{\partial y} \Big)$, $\rho$ and $\mu$ are the density and viscosity of the fluid medium respectively, $p$ represents the hydrodynamic pressure, $\ve{u}$ is the fluid velocity and $\mu[\nabla\ve{u} + (\nabla\ve{u})^\prime]$ is the viscous stress experienced by the fluid ($[\nabla\ve{u}]^\prime$ is the transpose of $\nabla\ve{u}$). The stress due to the NLC ($\ve{\sigma}$) is given by \cite{giomi2012banding,fukuda2004dynamics,marenduzzo2007hydrodynamics,sonnet2004continuum,Sengupta_PRL},
\begin{align} 
\label{eq.lc.stress}
\ve{\sigma} &= -\lambda s \ve{h} + \ve{qh - hq},
\end{align}
where, $s=\sqrt{2}\vert\ve{q}\vert$ is the scalar order parameter; $\ve{h}$ is the molecular field, which controls the relaxation to equilibrium and is given by
\begin{align}
 \label{eq.h}
\ve{h} &= \kappa \nabla^2 \ve{q} - A \ve{q} - C {|\ve{q}|}^2 \ve{q}.
\end{align}
Here, $\ve{q}$ is the nematic order parameter, a symmetric and a traceless second rank tensor having non-zero equal eigenvalues, used to describe the orientational order of the liquid crystals and referred to as the two-dimensional LdG tensor \cite{deGennes},
\begin{equation} \label{eq.Q}
\ve{q} =
	\begin{pmatrix}
		q_{11} & q_{12} \\
		q_{12} & -q_{11}
	\end{pmatrix};
\end{equation}
$\kappa$ is the elastic constant of the liquid crystals, $A$ and $C$ are material and temperature-dependent coefficients and $\lambda$ is the (dimensionless) NLC alignment parameter, which reflects whether the response of the NLC is affected by the fluid strain or vorticity and is determined experimentally~\cite{cates2009lattice,liverpool2006rheology,marenduzzo2007hydrodynamics}. 

The tensor $\ve{q}$ and the director $\ve{n}$ are related by $\ve{q} = s \left(\ve{n}\otimes \ve{n} - {I_2}/{2}\right)$ where $I_2$ is the identity matrix in 2D and $s^2 = 2 |\ve{q}|^2$. We can recover the director angle in the 2D system by the relation $\theta = \frac{1}{2}\tan^{-1}({q_{12}}/{q_{11}})$. 
We also think of $\ve{q}$ as being a 2D vector with two independent components, $\ve{q} = (q_{11}, q_{12})$ where 
$q_{11} = \frac{s}{2}\cos 2 \theta$ and $q_{12} = \frac{s}{2}\sin 2 \theta$.
In \cite{giomi2012banding}, the evolution equation for the $\ve{q}$-tensor is given by~\cite{beris_edwards_book,giomi2012banding,chen2017global}
\begin{equation} \label{eq.Q.tensor}
\frac{\partial \ve{q}}{\partial t} + \ve{u} \cdot \nabla \ve{q} = \frac{1}{2}\lambda s [\nabla\ve{u} + {(\nabla\ve{u})}^\prime] + \ve{q\omega - \omega q} + \frac{1}{\Gamma} \ve{h},
\end{equation}
where $\Gamma$ is the rotational diffusion coefficient \cite{marenduzzo2007hydrodynamics} and $\ve{\omega}= \nabla \times \ve{u}$ is the vorticity tensor. We impose strong homeotropic boundary conditions on the channel boundaries so that 
\begin{equation} \label{eq:q_bc}
\ve{q} = \bigg( \frac{2|A|}{C} \bigg) ^{1/2} \bigg( \ve{\nu}_\pm \otimes \ve{\nu}_\pm - \frac{\ve{I}_2}{2} \bigg), 
\end{equation}
where here $\ve{\nu}_\pm = \left(0, \pm 1 \right)$ correspond to the unit outward normals to the channel walls at $y = \pm L_2$, where $L_2$ is the channel half height, and $\ve{I}_2$ is the 2D identity matrix. We note that for homeotropic anchoring, $\ve{\nu}_\pm=\ve{n}$, the director at the walls.  

On the particle surface we apply a mixed anchoring condition,
\begin{align}
\label{eq:anchoring particle}
-\kappa \nabla q_{11}\cdot \ve{\nu}_p&=w\left(q_{11}+\sqrt{\frac{A}{2C}}\right),\\
-\kappa\nabla q_{12}\cdot \ve{\nu}_p&=wq_{12},
\end{align}
where $\ve{\nu}_p$ is the unit normal to the particle surface and $w$ is an anchoring strength parameter. When $w = 0$, we reduce to the Neumann boundary conditions, while $w \rightarrow \infty$ is the strong homeotropic anchoring limit.

\subsection{Non-dimensionalization}
\label{non-dimen}

We non-dimensionalize equations \eqref{eq.continuity}--\eqref{eq.Q.tensor} by applying the following scalings:
\begin{align}
\label{eq:non-dimensionalization}
X&= \frac{x}{L_2}, & Y&=\frac{y}{L_2}, & U&=\frac{u}{u_0}, \nonumber \\ 
V&=\frac{v}{u_0}, & P&=\frac{pL_2}{\mu u_0}, & T&=\frac{u_0 t}{L_2}, 
\end{align}
where $u_0$ is the mean channel velocity and $\mu$ is the viscosity of the fluid. The dimensionless versions of Eqs.~\eqref{eq.continuity}--\eqref{eq.h} are then 
\begin{align}
\label{eq.non.dim.cont}
\frac{\partial U}{\partial X} + \frac{\partial V}{\partial Y} = 0,
\end{align}
\begin{multline}
\label{eq.x-Re}
Re \bigg(\pdf{U}{T}+ U\frac{\partial U}{\partial X} + V\frac{\partial U}{\partial Y} \bigg) = -\frac{\partial P}{\partial X} +
\frac{\partial^2 U}{\partial X^2} + \frac{\partial^2 U}{\partial Y^2} \\ 
+ \frac{1}{Er} \frac{|A^*|}{2C^*} \frac{\partial}{\partial X}(-\lambda S H_{11}) + \frac{1}{Er}\frac{|A^*|}{2C^*} \frac{\partial}{\partial Y}(-\lambda S H_{12} - \eta),
\end{multline}
\begin{multline}
\label{eq.y-Re}
Re \bigg(\pdf{V}{T}+ U\frac{\partial V}{\partial X} + V\frac{\partial V}{\partial Y} \bigg) = -\frac{\partial P}{\partial Y} + \frac{\partial^2 V}{\partial X^2} + \frac{\partial^2 V}{\partial Y^2} \\
+ \frac{1}{Er} \frac{|A^*|}{2C^*} \frac{\partial}{\partial X}(-\lambda S H_{12} + \eta) + \frac{1}{Er} \frac{|A^*|}{2C^*} \frac{\partial}{\partial Y}(\lambda S H_{11}).
\end{multline}
Here $\ve{H}$ is the dimensionless molecular field described by
\begin{equation} \label{eq.H}
\ve{H} = \frac{\partial^2 \ve{Q}}{\partial X^2} + \frac{\partial^2 \ve{Q}}{\partial Y^2} - A^*\bigg(1 + \frac{1}{4} S^2 \bigg) \ve{Q},
\end{equation}
where $A^* = A L_2^2 / \kappa$, $C^* = CL_2^2/ \kappa$, $\ve{H} = \ve{h} \frac{L_2^2}{\kappa} \sqrt{\frac{2C^*}{|A^*|}}$, $\ve{Q} = \ve{q}\sqrt{\frac{2C^*}{|A^*|}}$, $\eta = 2({Q}_{12}{H}_{11} - {Q}_{11}{H}_{12})$, and $S=s\sqrt{\frac{2C^*}{|A^*|}} $; $Er = {u_0 \mu L_2}/{\kappa}$ denotes the Ericksen number, which represents the ratio of the viscous to elastic forces due to the NLC and $Re = {\rho u_0 L_2}/{\mu}$ is the Reynolds number, which indicates the relative magnitude of the inertial to viscous forces. In microfluidic flows, $Re \ll 1$ (Table \ref{table.params} gives typical operating regimes  $10^{-6} < Re < 10^{-3}$), which reduces Eqs.~\eqref{eq.x-Re}--\eqref{eq.y-Re} to a Stokes flow (where the inertial terms on the left-hand side are ignored), so that
\begin{multline} 
\label{eq.x-Re-small}
\frac{\partial P}{\partial X} =
\frac{\partial^2 U}{\partial X^2} + \frac{\partial^2 U}{\partial Y^2} + \frac{1}{Er} \frac{|A^*|}{2C^*} \frac{\partial}{\partial X}(-\lambda S H_{11}) \\ 
+ \frac{1}{Er}\frac{|A^*|}{2C^*} \frac{\partial}{\partial Y}(-\lambda S H_{12} - \eta),
\end{multline}
\begin{multline} 
\label{eq.y-Re-small}
\frac{\partial P}{\partial Y} = \frac{\partial^2 V}{\partial X^2} + \frac{\partial^2 V}{\partial Y^2} + \frac{1}{Er} \frac{|A^*|}{2C^*} \frac{\partial}{\partial X}(-\lambda S H_{12} + \eta) \\
+ \frac{1}{Er} \frac{|A^*|}{2C^*} \frac{\partial}{\partial Y}(\lambda S H_{11}).
\end{multline}

For the flow problem, we apply the following boundary conditions: no slip and no penetration on the channel walls and particle surface, 
\begin{align} \label{eq.bc.no-slip}
U&=0, & V&=0,
\end{align}
on $Y=\pm 1~\forall X$  and $X^2 + Y^2 = R^2$, where $R=r/L_2$ is the dimensionless radius of the colloidal particle and pressure boundary condition at the channel entrance and exit, 
\begin{align} \label{eq.bc.pr-inlet}
P&=1 &\textrm{on}\quad X&=-L_1/2 L_2, & -1&\leq Y \leq 1, \\
\label{eq.bc.pr-outlet}
P&=0 &\textrm{on}\quad X&=L_1/2 L_2, & -1&\leq Y \leq 1.
\end{align}

The dimensionless versions of the evolution equations~\eqref{eq.Q.tensor} are
\begin{multline} \label{eq.Q11.nondim}
\pdf{Q_{11}}{T}+U \frac{\partial Q_{11}}{\partial X} + V \frac{\partial Q_{11}}{\partial Y} = \lambda S \frac{\partial U}{\partial X}\\ - Q_{12} \bigg( \frac{\partial V}{\partial X} - \frac{\partial U}{\partial Y} \bigg) + \frac{\mu / \Gamma}{Er} H_{11},
\end{multline}
\begin{equation} \label{eq.Q12.nondim}
\begin{aligned}
\pdf{Q_{12}}{T}+U \frac{\partial Q_{12}}{\partial X} + V \frac{\partial Q_{12}}{\partial Y} = \frac{1}{2} \lambda S \bigg( \frac{\partial V}{\partial X} + \frac{\partial U}{\partial Y} \bigg) - \\
Q_{11} \bigg( \frac{\partial V}{\partial X} - \frac{\partial U}{\partial Y} \bigg) + \frac{\mu / \Gamma}{Er} H_{12}.
\end{aligned}
\end{equation}
Note that while the inertial terms may be ignored due to the smallness of $Re$, the corresponding terms in the $\ve{Q}$-tensor equation (the left-hand side of Eqs.~\ref{eq.Q11.nondim} and \ref{eq.Q12.nondim}) are order one and so cannot be neglected. At the channel walls we impose strong homeotropic boundary conditions which translate to (referring to Eq. \ref{eq:q_bc}),
\abeqn{eq.hom}{
Q_{11} = -1, \hspace{20mm}  Q_{12} = 0.
}

The anchoring conditions on the particle, \eqref{eq:anchoring particle}, become 
\begin{subequations} \label{eq.anch.var}
\begin{equation}
-\tilde{\nabla} Q_{11} \cdot \ve{\nu}_p = W \big( Q_{11} + 1 \big),
\end{equation}
\begin{equation}
-\tilde{\nabla} Q_{12} \cdot \ve{\nu}_p = W Q_{12},
\end{equation}
\end{subequations}
where $\tilde{\nabla}=(\partial/\partial X,\partial/\partial Y)$ is the dimensionless gradient operator, and $W=wL_2/ \kappa$ is the dimensionless anchoring parameter.

\subsection{Force exerted on the particle}
The total dimensional force (per unit length), $\ve{f}$, experienced by a particle of radius $r$, placed within the nematic fluid can be expressed as a line integration of the total stress acting on the particle~\cite{happel2012low, stark2002recent, stark2001stokes},
\begin{equation} \label{eq.force.nondim}
\ve{f} = \int_{\psi} \big( -p\ve{I} + \mu [\nabla\ve{u} + {(\nabla\ve{u})}^\prime] + \ve{\sigma} \big) \cdot \ve{\nu}_p \,\mathrm{d} \xi,
\end{equation}
where $\psi$ describes the perimeter of the circular particle. The displacement of the particle due to the force $\ve{f}$ can be calculated from the Stokes drag equation
\begin{equation} \label{eq.stokes.drag.particle}
\ve{f} = 3 \pi \mu (\ve{u} - \dot{\ve{x_p}}),
\end{equation}
where $\ve{x_p} = (x_p,y_p)$ denotes the instantaneous position of the centre of the particle and a dot represents differentiation with respect to time.

Non-dimensionalizing Eq.~\eqref{eq.force.nondim} via \eqref{eq:non-dimensionalization} and choosing the natural force scaling $\ve{F}=\ve{f}/\mu u_0$ gives the dimensionless drag ($F_x$) and lift ($F_y$) components of the force experienced by the particle ({i.e.}, the forces in the $x$ and $y$ directions, respectively),

\begin{multline}
\label{eq.fx}
F_x = \int_{\Psi} \bigg[ \bigg( 2 \frac{\partial U}{\partial X} - P - \frac{1}{Er} \frac{|A^*|}{2C^*} \lambda  S H_{11} \bigg)\nu_x \\
+ \bigg( \frac{\partial U}{\partial Y} + \frac{\partial V}{\partial X} - \frac{1}{Er} \frac{|A^*|}{2C^*} \big(\lambda S H_{12} + \eta \big) \bigg)\nu_y \bigg] \mathrm{d} \Omega,
\end{multline}
\begin{multline}
\label{eq.fy}
F_y = \int_{\Psi} \bigg[ \bigg( \frac{\partial U}{\partial Y} + \frac{\partial V}{\partial X} - \frac{1}{Er} \frac{|A^*|}{2C^*} \big( \lambda S H_{12} - \eta \big)\bigg)\nu_x \\
+ \bigg( 2 \frac{\partial V}{\partial Y} - P + \frac{1}{Er}\frac{|A^*|}{2C^*} \lambda S H_{11} \bigg)\nu_y \bigg] \mathrm{d} \Omega,
\end{multline}
where $\Psi$ describes the perimeter of the particle in the dimensionless domain and $\ve{\nu}_p=(\nu_x,\nu_y)$. The dimensionless version of \eqref{eq.stokes.drag.particle} reads as 
\begin{subequations} \label{eq.pos}
	\begin{equation} 
		3 \pi (U - \dot{X}_P ) = F_x,
	\end{equation}
	\begin{equation} 
		3 \pi (V - \dot{Y}_P ) = F_y,
	\end{equation}
\end{subequations}
where $\dot{X}_P = \mathrm{d} X_P/ \mathrm{d} T$ and $\dot{Y}_P = \mathrm{d} Y_P/ \mathrm{d} T$ are the respective dimensionless particle velocity components.

\begin{table} [!ht]
\caption{Typical values of the physical parameters}
\begin{tabular}{ l r }
\hline \\[-2mm]
Parameter & Typical values \\
$[$units$]$ & $[$reference$]$ \\ [1mm]
\hline \\[-2mm]
Elastic constant, $\kappa$ [$pN$] & 40 \cite{Sengupta_PRL}\\
Length of the microchannel, $L_1$ [$\mu m$] & 50 \\
Half-height of the microchannel, $L_2$ [$\mu m$] & 10 \\
Particle radius, $r$ [$\mu m$] & 3 \\
Mean fluid velocity, $u_0$ [$\mu m/s$] & 10\\
Rotational diffusion constant, $\Gamma$ [$Pa\,s$] & 7.3 \cite{Sengupta_PRL}\\
Viscosity, $\mu$ [$Pa\,s$] & 0.01 \cite{anderson2015transitions}\\
NLC material property, $A$ [$MJ/m^3$] & -0.172 \cite{Sengupta_PRL} \\
NLC material property, $C$ [$MJ/m^3$] & 1.72 \cite{Sengupta_PRL} \\ [1mm]
\hline \\[-2mm]
\multicolumn{2}{l}{Dimensionless parameters used in calculation} \\[1mm]
\hline \\[-2mm]
NLC alignment parameter, $\lambda$ & 1 \cite{Sengupta_PRL} \\
Dimensionless particle radius, $R$ & 0.3 \\
Reynolds number, $Re$ & 0.0001 \\
Ericksen number, $Er$ & 0.001 -- 10 \\
Parameter, $|A^*|/C^*$ & 0.1 \\
Relative anchoring strength, $\log W$ & 3 \\ [1mm]
\hline
\end{tabular}
\label{table.params}
\end{table}

\subsection{Computational details}
The numerical computations are carried out using the finite-element-based commercial package COMSOL~v5.2~\cite{pryor2009multiphysics, siegel2008review, cai2011mathematical, li2009comsol, van2009integrated}. The discretization of the variational form of the governing equations follows the standard Galerkin formalism \cite{hu2001direct}. To save computational cost, the Jacobian matrix is updated once in several iterations. Within each Newton iteration, the sparse linear system is solved by preconditioned Krylov methods \cite{elman2014finite} such as the generalized minimum residual (GMRES) method \cite{saad1986gmres, saad1993flexible} and the bi-conjugate gradient stabilized (BCGSTAB) method \cite{barrett1994templates}. The weak boundary condition \eqref{eq.anch.var} is solved using an inexact Newton method with backtracking for enhanced convergence and stability \cite{eisenstat1996choosing}. The segregated solver \cite{pryor2009multiphysics} is used with a relative tolerance of 0.001. Two segregated steps decoupling the hydrodynamics and the director equations are set up. The solution domain is meshed with free triangular elements (first shape order) and around 180,000 degrees of freedom are solved for. Boundary layer meshing of five layers is constructed near the no-slip boundaries. 

For the particle dynamics calculation, the time-dependent solver is utilized with a relative tolerance of $10^{-4}$ along with the moving mesh method based on Arbitrary Lagrangian--Eulerian (ALE) formulation \citep{donea1982arbitrary, hu2001direct, anderson2004arbitrary}, which enables the moving boundaries without the need for mesh movement. This allows for mesh deformation with the translation of the physical boundary. Smoothing the deformation (because of the movement of the interior nodes) throughout the domain is achieved by solving the hyperelastic \cite{yamada1993arbitrary, curtis2013axisymmetric} smoothing partial differential equation with the boundary conditions specified by the particle boundary movement. Second-order shape elements are used to mesh the geometry in the case of free deformation with particle movement. The spatial derivatives of the dependent variables are now transformed with respect to the new transformed coordinates. The displacement of the particle boundary is related to their respective velocities ($\dot{X}_P, \dot{Y}_P$). When the mesh deformation produces a low-element quality (defined as the ratio of the inscribed to circumscribed circles’ radii for each simplex element), where the minimum element quality is less than 0.3 (it is 0 for the degenerated element and 1 for the best possible element), remeshing of the entire domain is done and the simulation progresses. 

\section{Results and Discussion}
\subsection{Static particle and no fluid flow} \label{section__eq}

\begin{figure}[!ht]
\begin{center}

\includegraphics[width=0.49\textwidth]{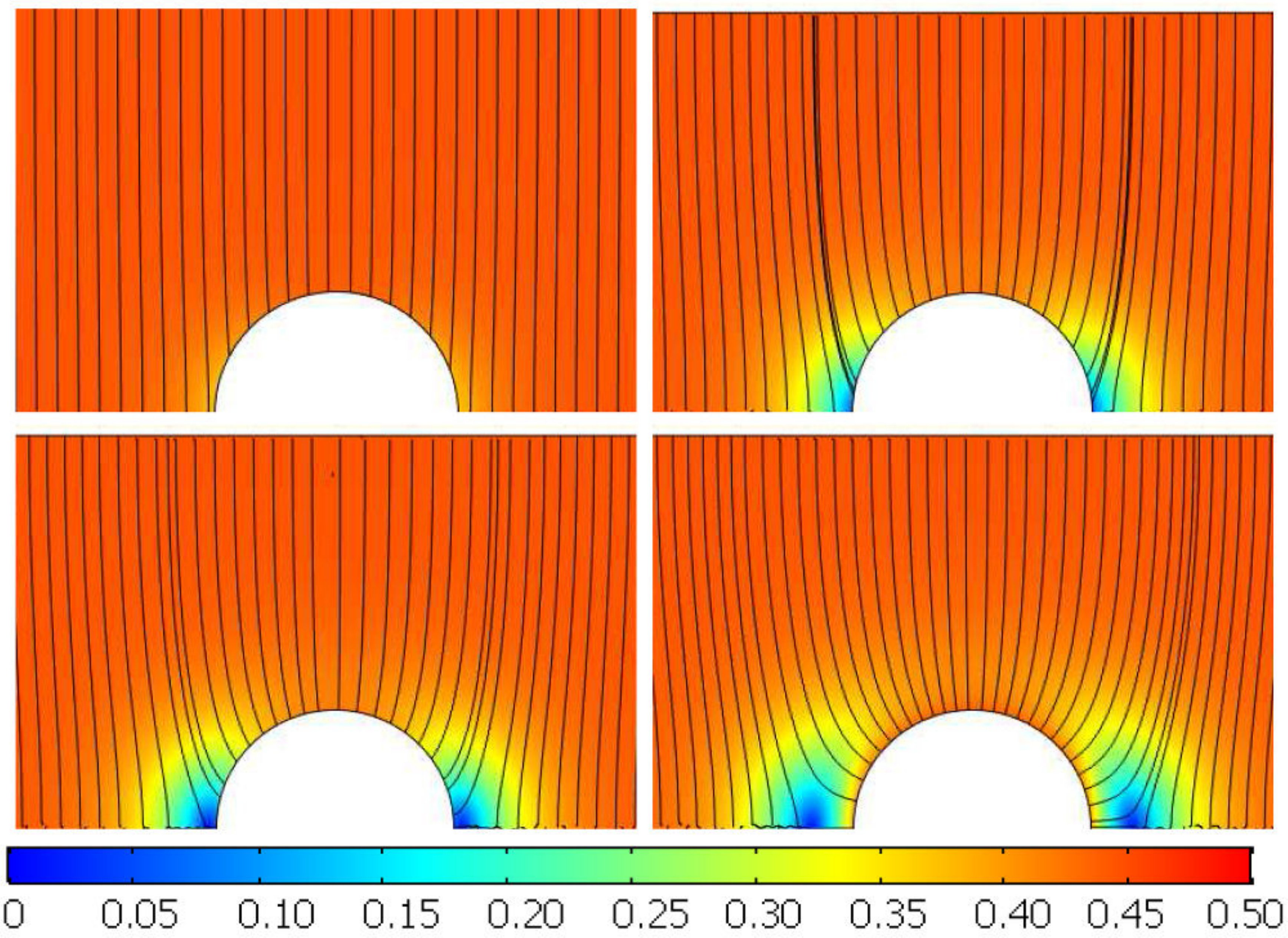}

\caption[.]{Orientation of the director field around the particle with homeotropic boundary conditions on the channel walls with zero flow field ($Er=0$) for $R=0.3$. The angle of the director orientation, $\theta$ (quiver angle) is given by $\theta = \frac{1}{2}\tan^{-1}({Q_{12}}/{Q_{11}})$. The anchoring strength on the particle surface is varied from weak to strong. (a) $\log W = 0$, (b) $\log W = 0.8$, (c) $\log W = 1$ and (d) $\log W = 2$. The solid lines represent the director profiles ($\theta$) and the background contour represent the magnitude of the order parameter $S$. Due to symmetry we show only half of the channel. The values of the relevant parameters used for the calculation are given in Table~\ref{table.params}.}
\label{fig.director.contour}
\end{center}
\end{figure}

The equilibrium director profiles and the order parameter \emph{S} are first calculated in the absence of a flow field $U=V=0$ (equivalently $Er=0$). In this case, we only need to solve Eqs. \eqref{eq.Q11.nondim} and \eqref{eq.Q12.nondim}, which reduce respectively to $H_{11} = H_{12} = 0$. These are solved subject to the boundary conditions in Eqs. \eqref{eq.hom} and \eqref{eq.anch.var}. We vary the anchoring strength at the particle surface from weak to strong (homeotropic) anchoring (Fig.~\ref{fig.director.contour}). Surface anchoring strengths can in practice be altered by photo-excitation \cite{Ikeda2003}, electric or magnetic fields \cite{deuling1972,schadt1971} or chemical surface functionalization \cite{jerome1991}. Defects near the boundary along the axial symmetry line ($Y=0$, both on the east and west side) are observed, which is in line with the literature reports \cite{kuksenok1996, Fukuda2006, toth2002hydrodynamics, ruhwandl1997monte}. The contours of the order parameter are also qualitatively similar to the report of Fukuda et al.~\cite{fukuda2004dynamics} and Sengupta et al.~\cite{Sengupta_review}.

\begin{figure}[!ht]
\begin{overpic}[width=0.45\textwidth]{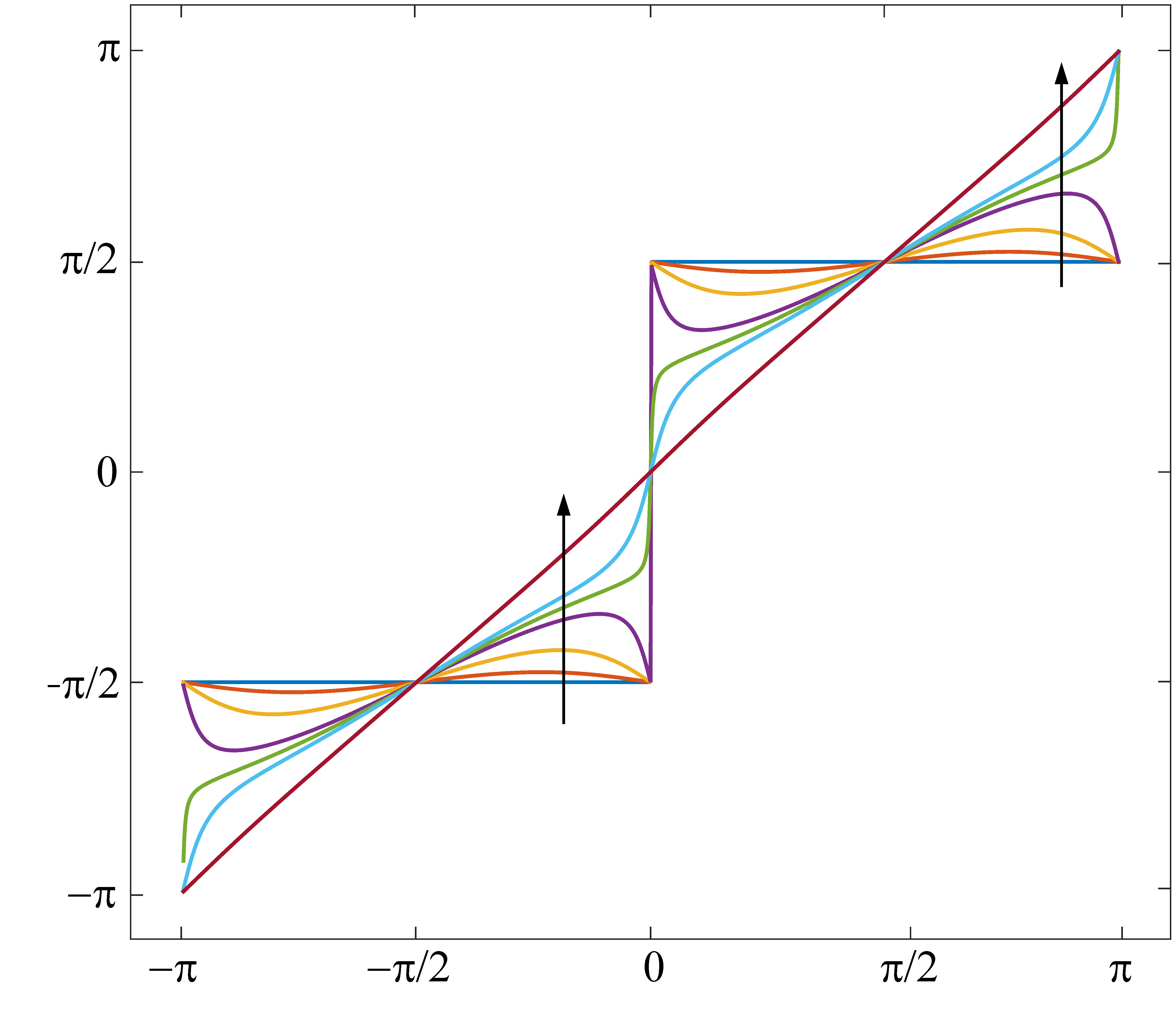}
\footnotesize
\put(40,0){Angular position ($\phi$)}
\put(-1,25){\rotatebox{90} {Director orientation angle ($\theta$)}}
\put(26,45){Increasing $W$}
\put(66,80){Increasing $W$}
\end{overpic}
\caption{Variation of the director angle on the particle surface
with angular location as the anchoring strength is increased
with $\log W = -3, 0, 0.5, 0.8, 0.9, 1$ and $2$, with zero flow field
($Er = 0$) for $R = 0.3$. The particle is located at the centre
of the channel with homeotropic boundary conditions. The
values taken for all other parameters used for the calculation
are given in Table~\ref{table.params}.}
\label{fig.director.anch.strength}

\end{figure}
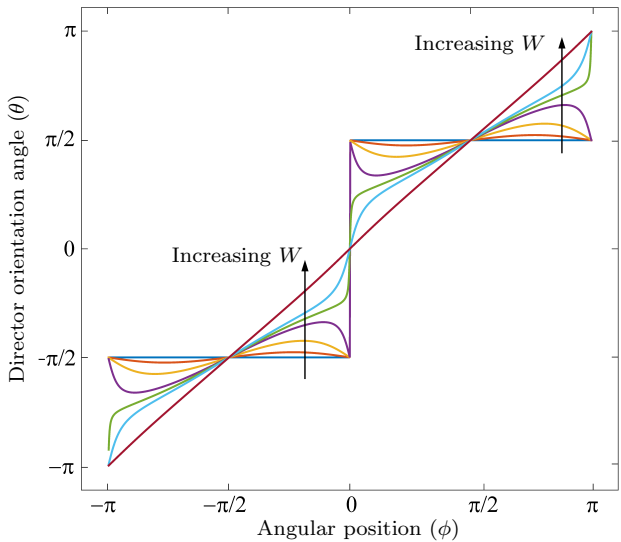

The director field inclination at the particle surface is very susceptible to the magnitude of $W$ in the range of $0.5 < \log W < 1$ for $R=0.3$ (Fig.~\ref{fig.director.anch.strength}). Within this range the anchoring switches from being effectively uniform (zero anchoring) to homeotropic (strong anchoring). Symmetry enforces that $\theta \big( \phi - \frac{\pi}{2} \big) = -\theta \big( - \phi - \frac{\pi}{2} \big)$ for $-\pi \leq \phi \leq 0$ and $\theta \big( \phi + \frac{\pi}{2} \big) = -\theta \big( - \phi + \frac{\pi}{2} \big)$ for $0 \leq \phi \leq \pi$.

\begin{figure}[!htbp]
\begin{overpic}[width=0.4\textwidth]{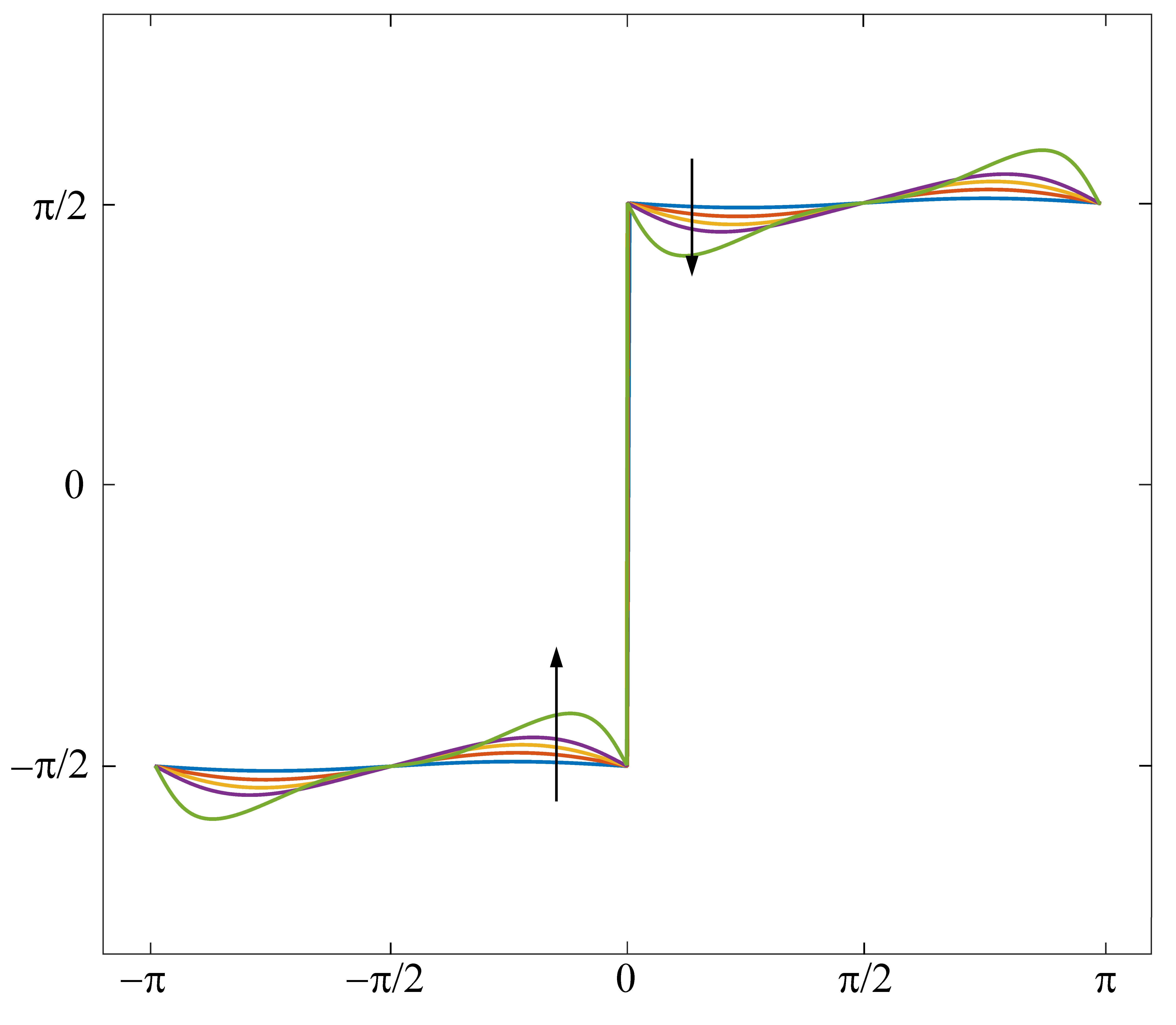}
\footnotesize
\put(15,75){(a)}
\put(40,-2){Angular position ($\phi$)}
\put(-3,20){\rotatebox{90} {Director orientation angle ($\theta$)}}
\put(23,32){Increasing $R$}
\end{overpic}

\vspace{4mm}

\begin{overpic}[width=0.4\textwidth]{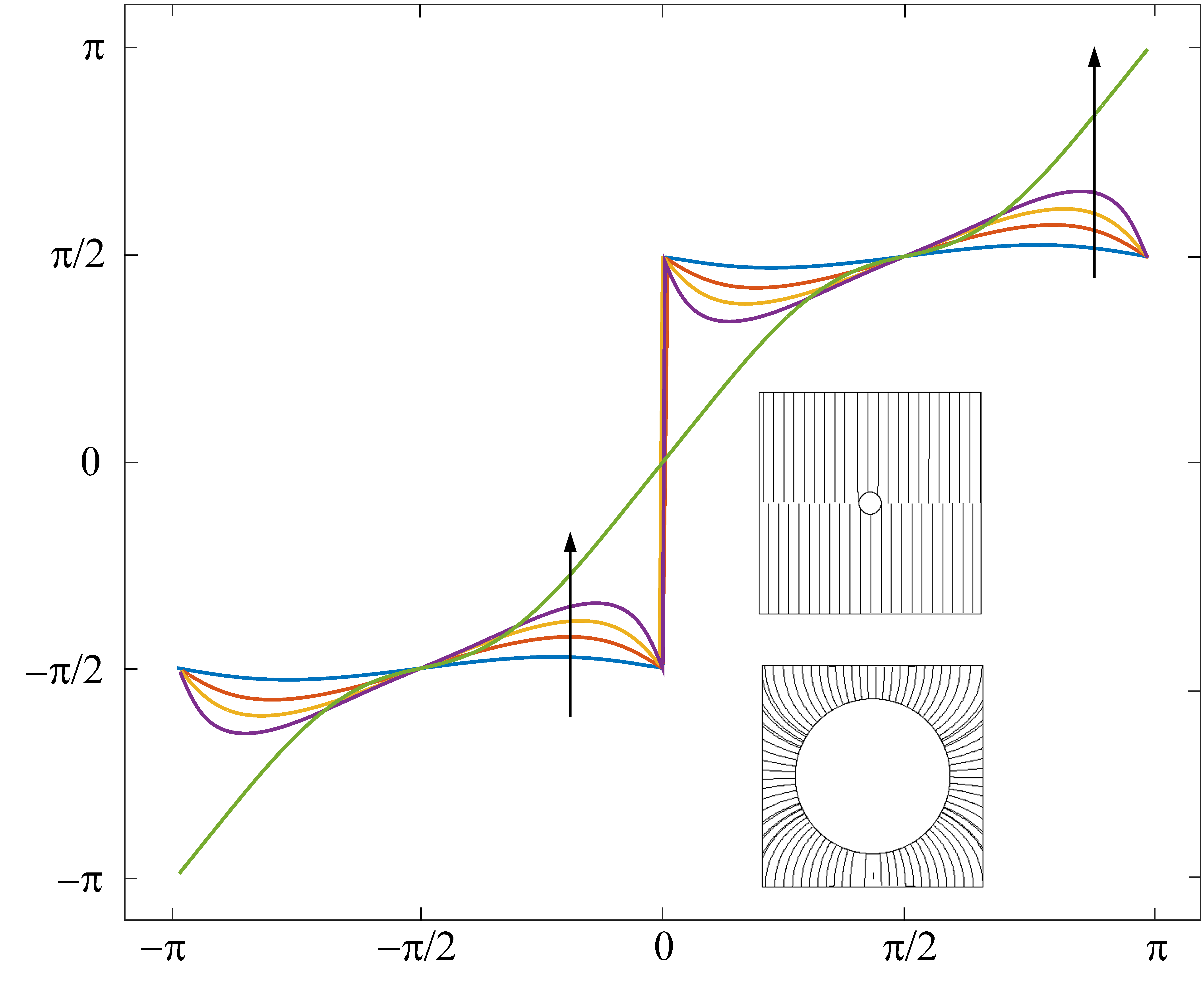}
\footnotesize
\put(17,72){(b)}
\put(40,-1){Angular position ($\phi$)}
\put(-3,20){\rotatebox{90} {Director orientation angle ($\theta$)}}
\put(84,9){\rotatebox{90} {\parbox{0.17\textwidth}{Director profiles when the relative size of the particle is increased}}}
\put(23,38){Increasing $R$}
\end{overpic}

\vspace{4mm}

\begin{overpic}[width=0.4\textwidth]{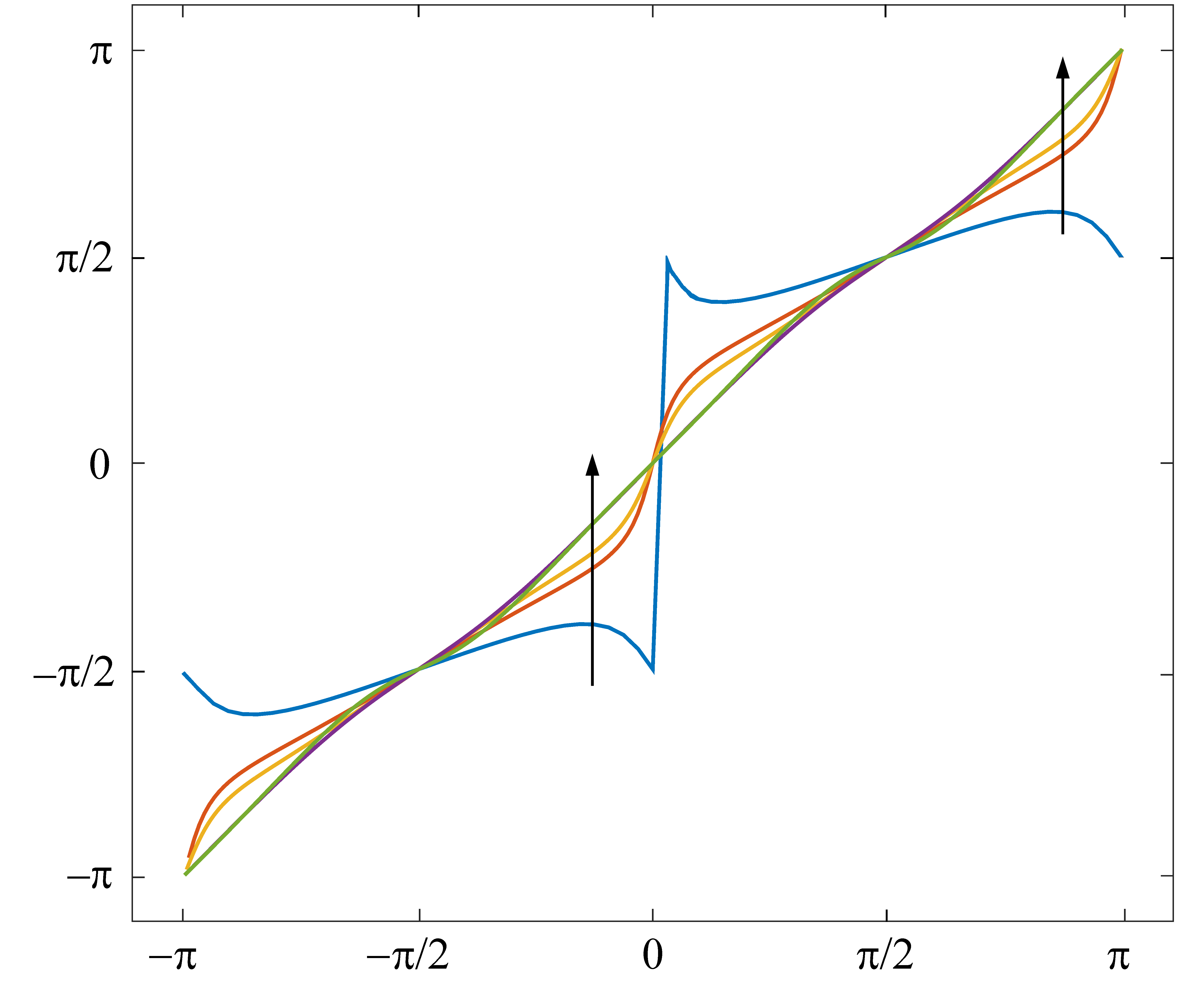}
\footnotesize
\put(18,73){(c)}
\put(40,-1){Angular position ($\phi$)}
\put(-3,20){\rotatebox{90} {Director orientation angle ($\theta$)}}
\put(26,46){Increasing $R$}
\end{overpic}

\caption[.]{Variation of the director angle on the particle surface, with zero flow field ($Er=0$) with different particle sizes and anchoring strength, (a) $\log W = 0$, (b) $\log W = 0.5$ (inset: qualitative visualization of the director orientation corresponding to $R=0.1$ and $0.8$), and (c) $\log W = 1$. Note that the particle is located at the centre of the channel with homeotropic boundary conditions. The values taken for all other parameters used for the calculation are given in Table~\ref{table.params}. The arrow represents the particle size in increasing order, as $R = 0.1, 0.3, 0.5, 0.7$ and $0.8$.}
\label{fig.director.part.size}
\end{figure}
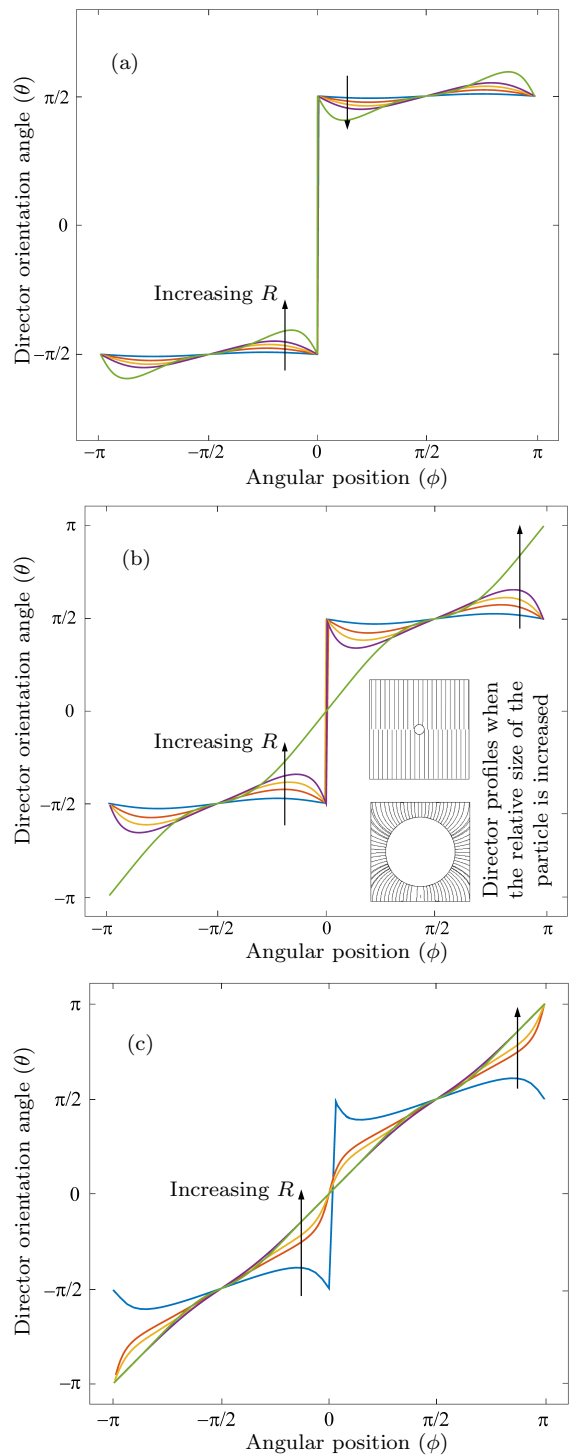

The effect of the particle size relative to the channel dimension has a profound influence on the director orientation (Fig.~\ref{fig.director.part.size}). The impact of the particle size is most pronounced in the case of strong anchoring (increasing $W$). This is because, with increasing particle size, the distance between the particle surface and channel walls reduces, which induces strong coupling between the directors on the particle surface and channel walls. As in the situation observed in Fig.~\ref{fig.director.anch.strength}, there is a narrow range of $R$ over which the director orientation switches from uniform to homeotropic on the particle boundary. The values of $R$ in this transition region decrease with increasing $W$ (Fig. \ref{fig.director.part.size}). For example, in the case of $W=3.2$ the transition occurs in between $0.7<R<0.8$ (Fig. \ref{fig.director.part.size}b), whereas for $W=10$, this occurs at $0.1<R<0.3$ (Fig. \ref{fig.director.part.size}c).

\subsection{Static particle with fluid flow effects}  \label{section__fluid_flow}

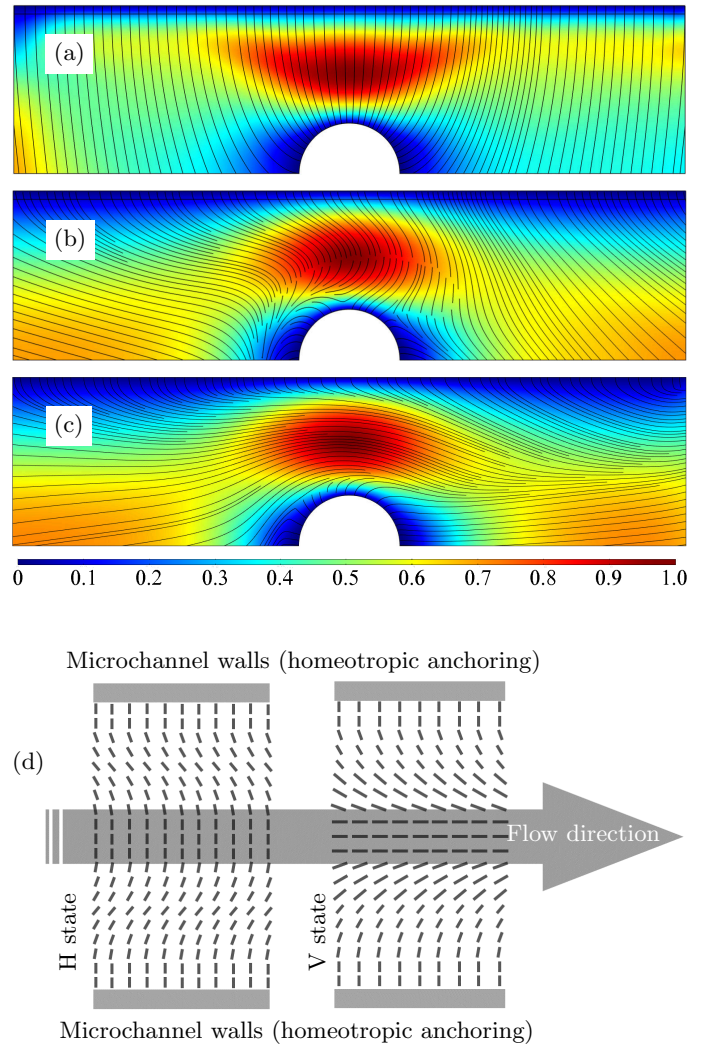
\begin{figure}[!ht]
\begin{overpic}[width=0.5\textwidth]{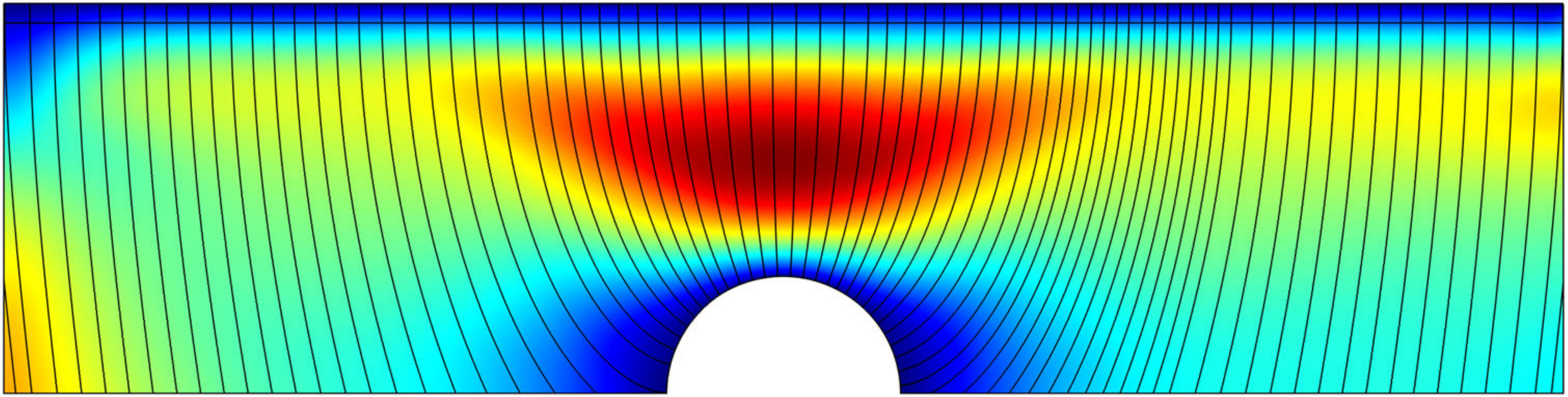}
\put(5,17){\colorbox{white}{(a)}}
\end{overpic}

\vspace{2mm}

\begin{overpic}[width=0.5\textwidth]{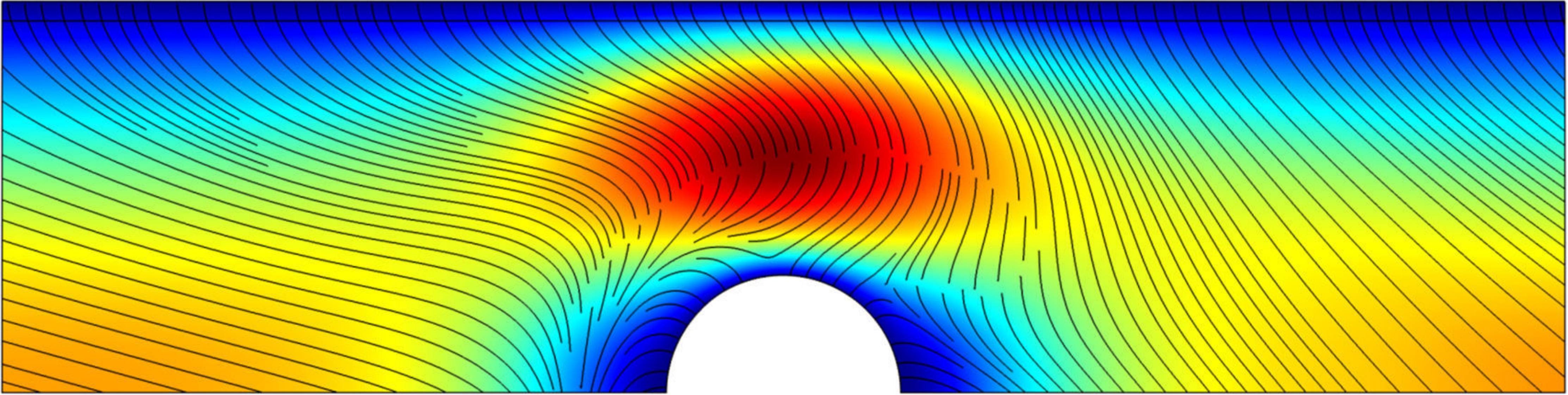}
\put(5,17){\colorbox{white}{(b)}}
\end{overpic}

\vspace{2mm}

\begin{overpic}[width=0.5\textwidth]{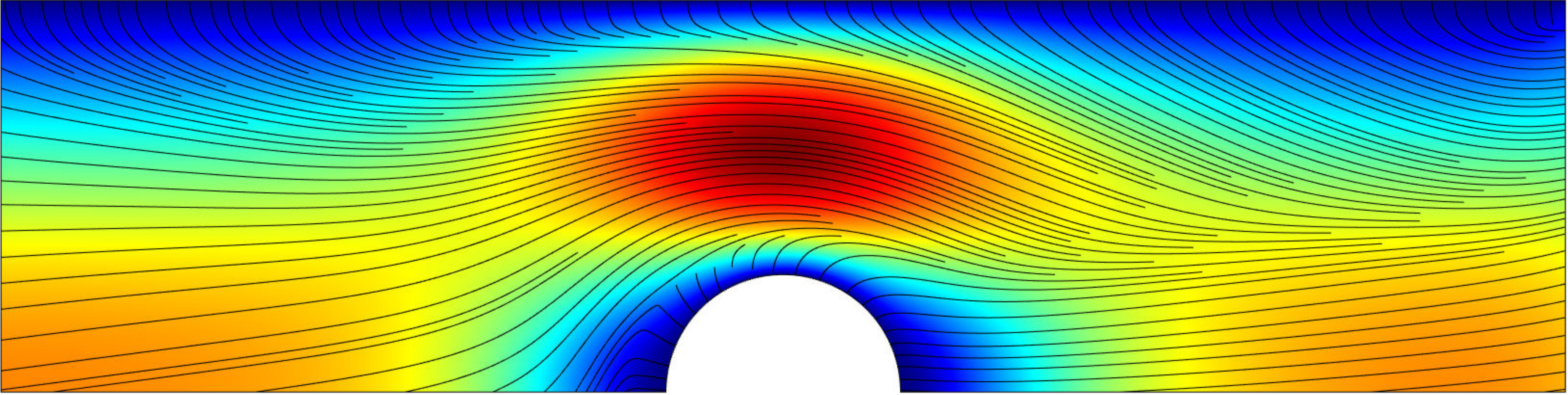}
\put(5,17){\colorbox{white}{(c)}}
\end{overpic}

\vspace{2mm}

\begin{overpic}[width=0.5\textwidth]{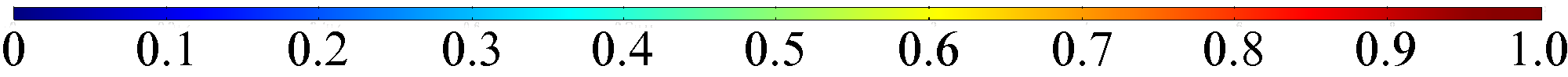}
\end{overpic}

\vspace{2mm}

\begin{overpic}[width=0.5\textwidth]{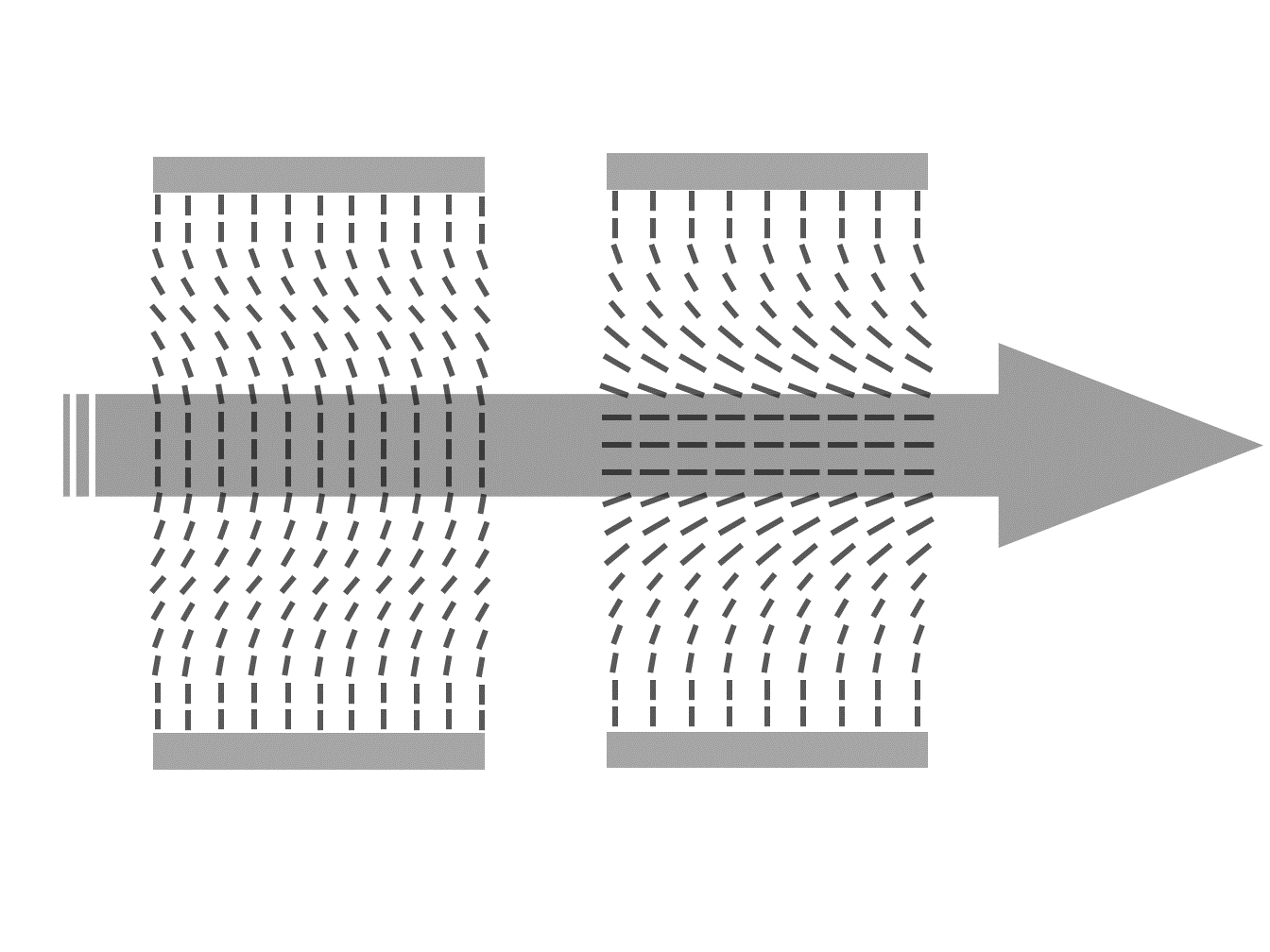}
\put(73.3,39.3){\color{white}{Flow direction}}
\put(0,50){(d)}
\put(44,20){\rotatebox{90}{V state}} 
\put(7,20){\rotatebox{90}{H state}} 
\put(7,10){Microchannel walls (homeotropic anchoring)}
\put(8,65){Microchannel walls (homeotropic anchoring)}
\end{overpic}
\caption[.]{Profiles of the director orientation for (a) $Er = 0.001$ (weak flow), (b) $Er = 0.1$ (intermediate flow), (c) $Er = 10$  (strong flow). Due to symmetry we show only half of the channel. (d) V and H state configurations of the director alignment \cite{jewell2009flow}. The background colour contour represents the axial velocity field. Here $R = 0.3$. The bottom edge of the domain is the symmetry condition. Homeotropic anchoring conditions are maintained on the channel walls as well as on the particle surface ($W \rightarrow \infty$). The values taken for all other parameters used for the calculation are given in Table~\ref{table.params}. }
\label{fig.flow.contour}
\end{figure}

We now consider the case where we include a flow field within the microchannel but hold the particle in place. Here we solve the Eqs. (\ref{eq.x-Re-small}), \eqref{eq.y-Re-small}, \eqref{eq.Q11.nondim}, \eqref{eq.Q12.nondim} subject to Eqs. (\ref{eq.bc.no-slip})--\eqref{eq.bc.pr-outlet}, \eqref{eq.hom}, \eqref{eq.anch.var}. We consider the steady-state situation for the director field, and so set ${\partial \ve{Q}}/{\partial T} = 0$ in Eq. \eqref{eq.Q11.nondim} and \eqref{eq.Q12.nondim}. This corresponds to a situation before a particle is allowed to move and also resembles a system with an obstacle. We increase the Ericksen number by increasing the strength of the flow field ($u_0$) (maintaining the condition of $Re \ll 1$) and observe three distinct regimes: weak, moderate and strong. For small $Er$, the director profiles remain almost undistorted compared to the previous section. Here the nematic forces dominate over the viscous forces and the flow profile significantly deviates from Poiseuille flow (Fig.~\ref{fig.flow.contour}a).

There are two stable topologically distinct states for the director field, known as the horizontal (H) and vertical (V) states (Fig.~\ref{fig.flow.contour}d), based on the director orientation profile. The H and V states have different orientations at the channel centre (H-state, $\theta = 2n\pi$ and in V-state, $\theta = (n+0.5)\pi$ where $n = 0,1,2,3...$). In the H-state the directors splay, whereas in the V-state they tend to bend \citep{towler2004liquid}. In the moderate flow regime (Fig.~\ref{fig.flow.contour}b), one primarily observes the V-state and the flow begins to assume a parabolic profile (Fig.~\ref{fig.flow.contour}b). As we increase the flow rate ({i.e.}, increase $Er$) further, the effect of the NLC stress on the hydrodynamics weakens (see Eqs. \ref{eq.x-Re-small} and \ref{eq.y-Re-small}). The director field transitions from the V-state to the more energetically favourable H-state configuration (Fig.~\ref{fig.flow.contour}c,d) and the flow assumes a fully developed Poiseuille profile. (Fig.~\ref{fig.flow.contour}c) \citep{jewell2009flow}. This is in line with the experimental observations of Sengupta et al.~\cite{Sengupta_PRL}.

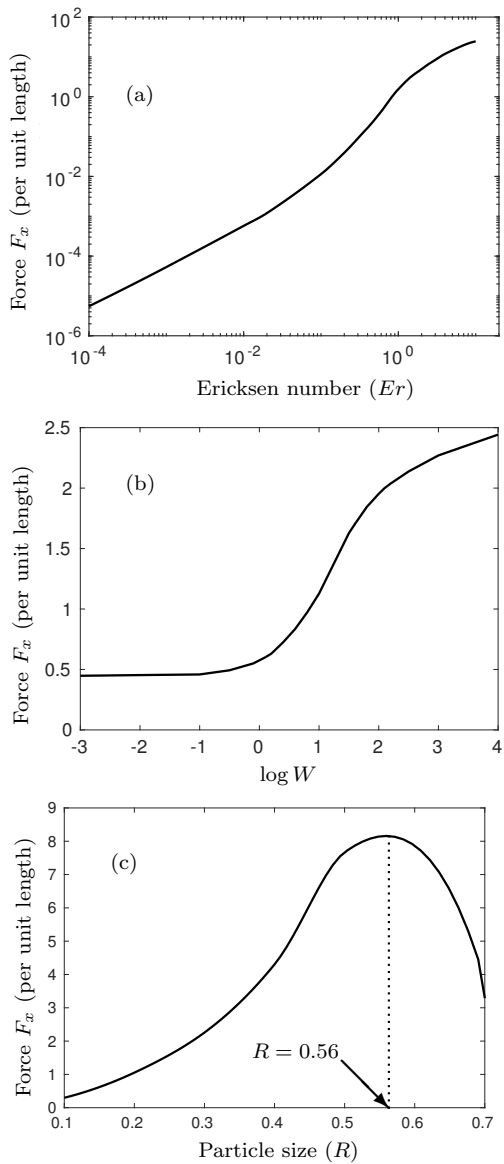
\begin{figure}[!htbp]
\begin{overpic}[width=0.36\textwidth]{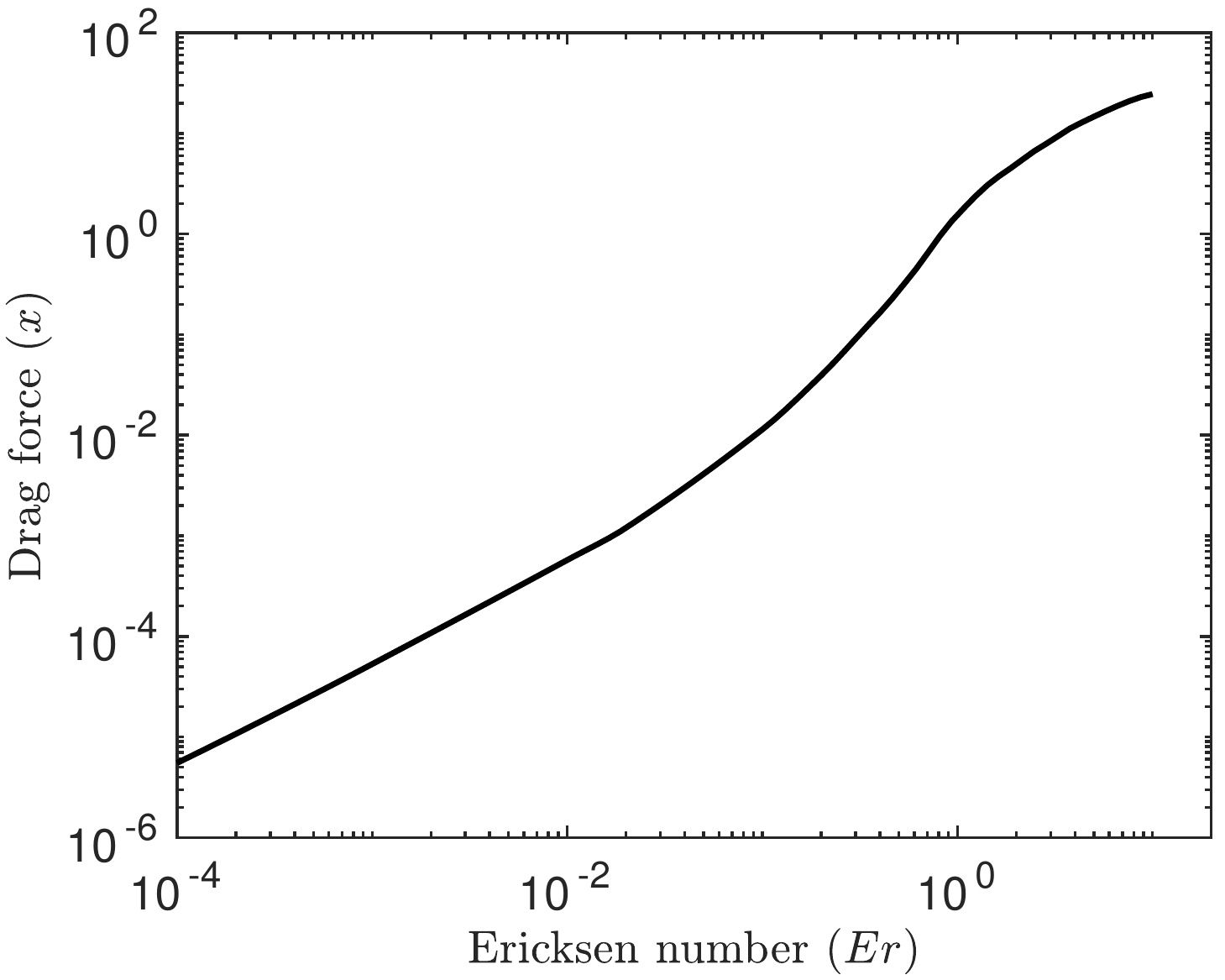}
\centering
\footnotesize
\put(35,-0.5){\colorbox{white}{Ericksen number ($Er$)}}
\put(-3.5,15){\rotatebox{90}{\colorbox{white}{Force $F_x$ (per unit length)}}}
\put(22,60){(a)}
\end{overpic}

\vspace{3mm}

\begin{overpic}[width=0.36\textwidth]{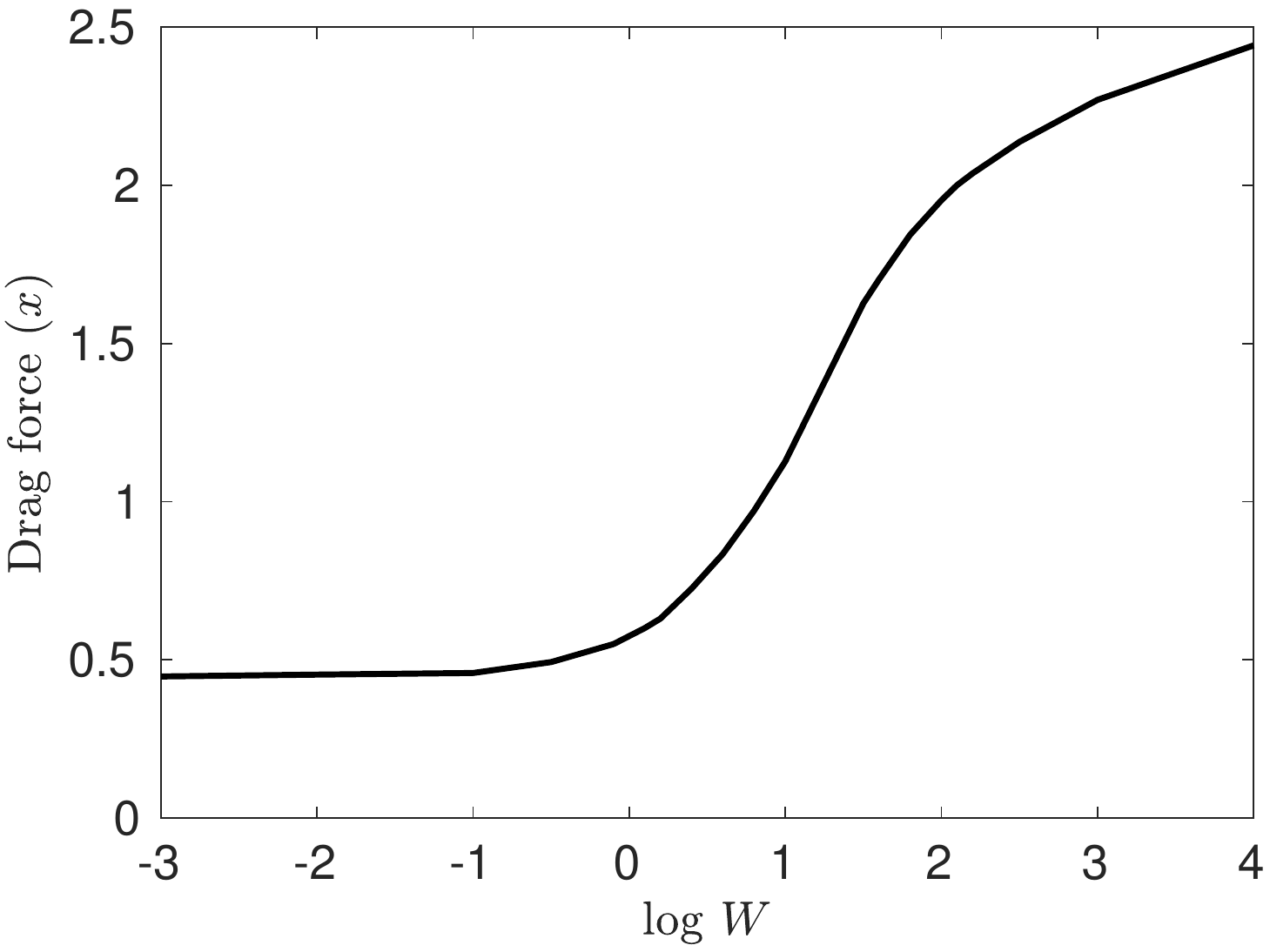}
\centering
\footnotesize
\put(48,0){\colorbox{white}{$\log W$}}
\put(-3,10){\rotatebox{90}{\colorbox{white}{Force $F_x$ (per unit length)}}}
\put(22,60){(b)}
\end{overpic}

\vspace{2mm}

\begin{overpic}[width=0.35\textwidth]{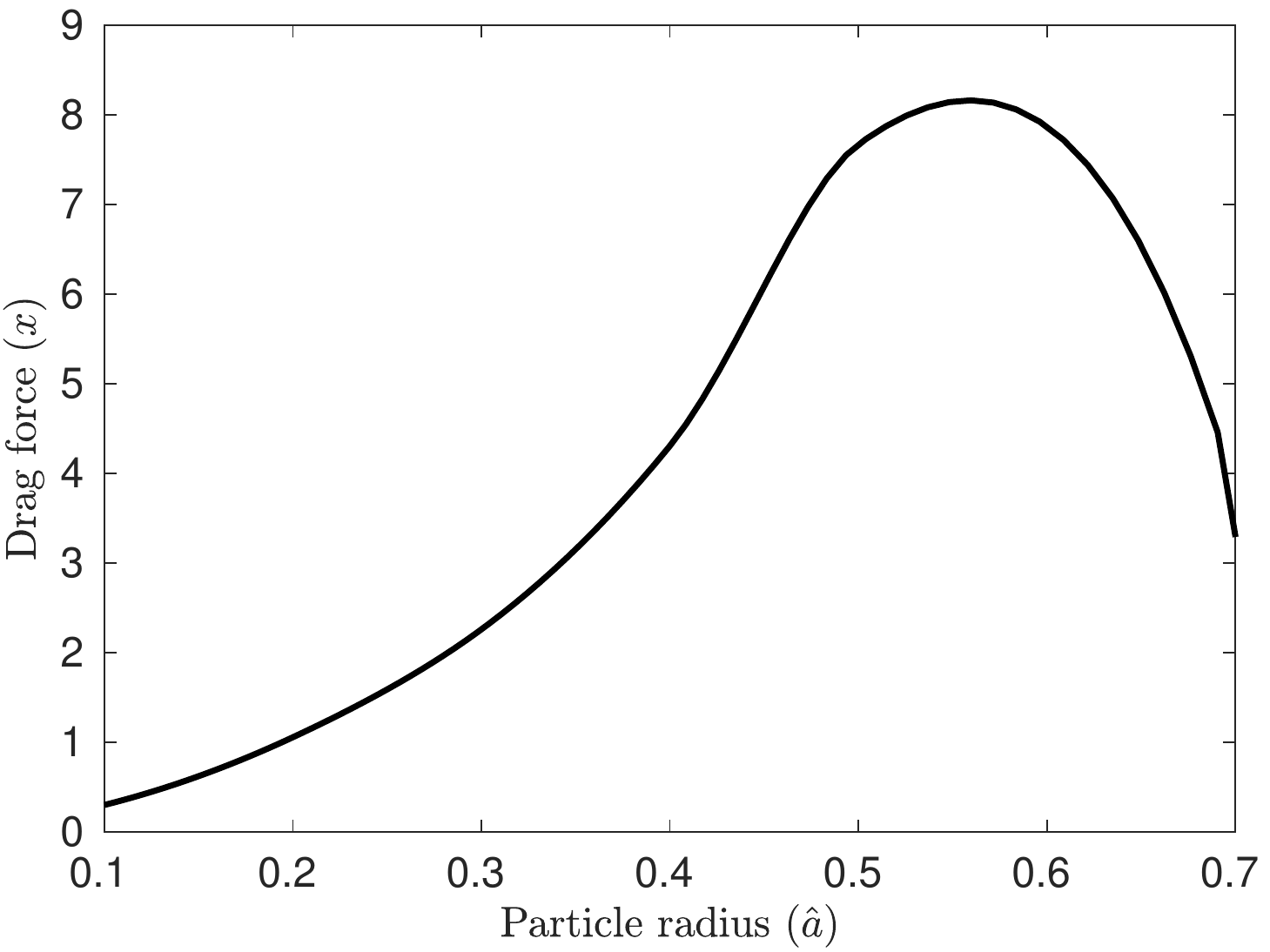}
\centering
\footnotesize
\thicklines
\put(35,-1){\colorbox{white}{Particle size ($R$)}}
\put(-4,10){\rotatebox{90}{\colorbox{white}{Force $F_x$ (per unit length)}}}
\put(18,60){(c)}
\dottedline{1.5}(77,9.5)(77,67)
\put(67,19.5){\vector(1,-1){10}}
\put(48,20){$R = 0.56$}
\end{overpic}
\caption[.]{The axial force $F_x$ experienced on a static particle as a function of the (a) Ericksen number ($Er$); (b) particle surface anchoring strength ($\log W$); and (c) particle size ($R$). The reference values of the parameters are $R = 0.3$, $Er = 1$ and $\log W = 3$. The values for all other parameters used for the calculation are given in Table~\ref{table.params}.}
\label{fig.forces}
\end{figure}

Of principal interest is the force experienced by a particle that is placed into the microchannel as a result of the viscous and elastic stresses exerted on the particle surface, given by Eq.~\eqref{eq.force.nondim}, since this ultimately dictates the motion of the particles within the flow. Since the particle is placed at the centre of the channel, the overall lift force ($F_y$) is zero due to symmetry and the only force experienced is in the $x$ direction, $F_x$ (drag). As the flow field increases (through increasing $Er$), the force exerted on the particle increases, as expected, in the direction of the hydrodynamic pressure gradient (Fig.~\ref{fig.forces}a). 

When the anchoring strength on the particle is increased the driving force experienced by the particle increases (Fig.~\ref{fig.forces}b). This suggests that tuning the particle surface anchoring conditions by external stimuli, such as by photo-excitation \cite{Ikeda2003} or an electric field \cite{deuling1972} could result in acceleration of the particle, leading to spatial reorganization in the nematohydrodynamic field.

An interesting observation is made on the dependence of the force on the particle size. The viscous force on a particle due to the hydrodynamic flow increases proportionally with particle radius \cite{leal2007advanced}. However, as the particle size is increased the effect of attractive normal boundary conditions on the channel walls is felt i.e., the director field is normal to both the channel surfaces and the particle boundary for large $W$ and this induces an attraction of the particle towards the channel walls, which resists the viscous stresses in the $x$ direction. As a result, a critical particle size exists for which the force is maximum ($R \approx 0.56$ for the parameters considered in Fig.~\ref{fig.forces}c). Coupled with the observations on particle force variations with anchoring strength and flow rate we see that the use of NLC could offer potential mechanisms for enhancing or impeding the transport of different cargo within a microchannel. Controlling the resistance to the motion of certain particles could also provide a new mechanism for particle self-assembly.

\newpage
\subsection{Particle dynamics with fluid flow}  \label{section__dyna}

\subsubsection{Single particle}

We now study the director profiles and particle trajectories when we allow the particle to move in response to the nematohydrodynamic field. Here we consider a specific Ericksen number ($Er = 0.02$) where the NLC elastic forces dominate over viscous forces  \cite{ruhwandl1995friction,fukuda2004dynamics,Sengupta_review}. In this case we must solve the transient set of Eqs. \eqref{eq.Q11.nondim}--\eqref{eq.anch.var} together with the flow hydrodynamics in Eqs. \eqref{eq.non.dim.cont}, \eqref{eq.x-Re-small}--\eqref{eq.bc.pr-outlet}. The particles are initially considered to be stationary ($\dot{X}_P = \dot{Y}_P = 0$) and released into the flow, and the initial condition for the NLC and the fluid flow is the equilibrium configuration assumed when keeping the particle position fixed (as found in Section~\ref{section__fluid_flow}). Also, we impose strong anchoring conditions on the particle surface. To compare this situation with a Newtonian fluid we simply set the elastic constant, $\kappa = 0$ (thereby $Er \rightarrow \infty$). 

\begin{figure}[!ht]
\includegraphics[width=0.45\textwidth]{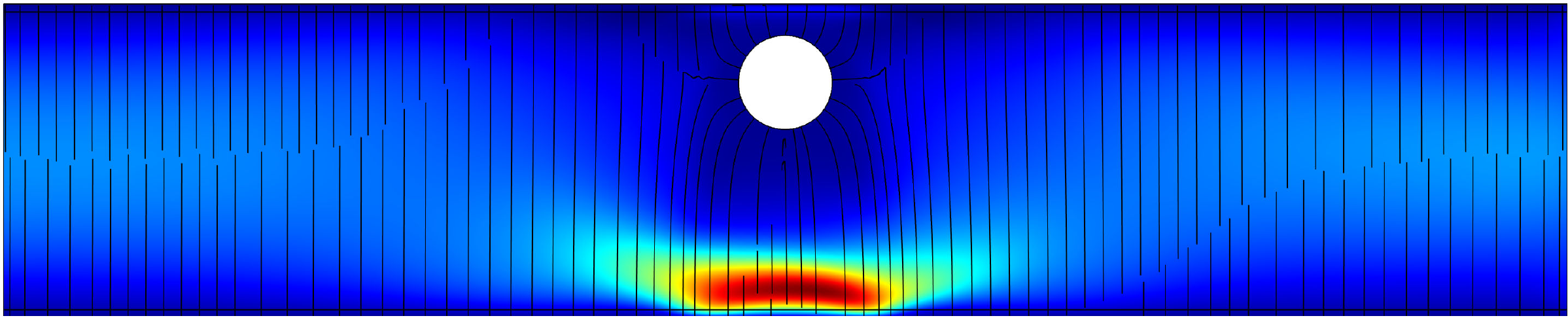}

\vspace{1mm}

\includegraphics[width=0.45\textwidth]{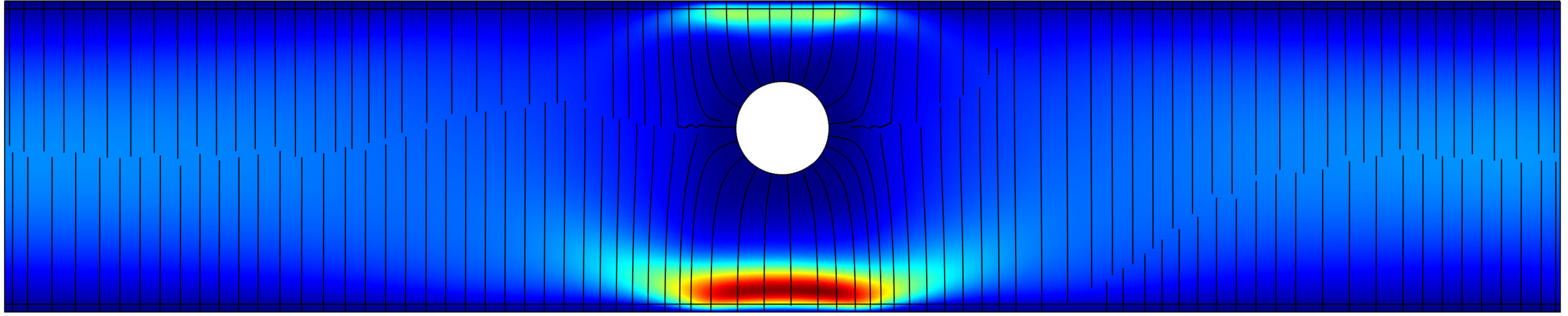}

\vspace{1mm}

\includegraphics[width=0.45\textwidth]{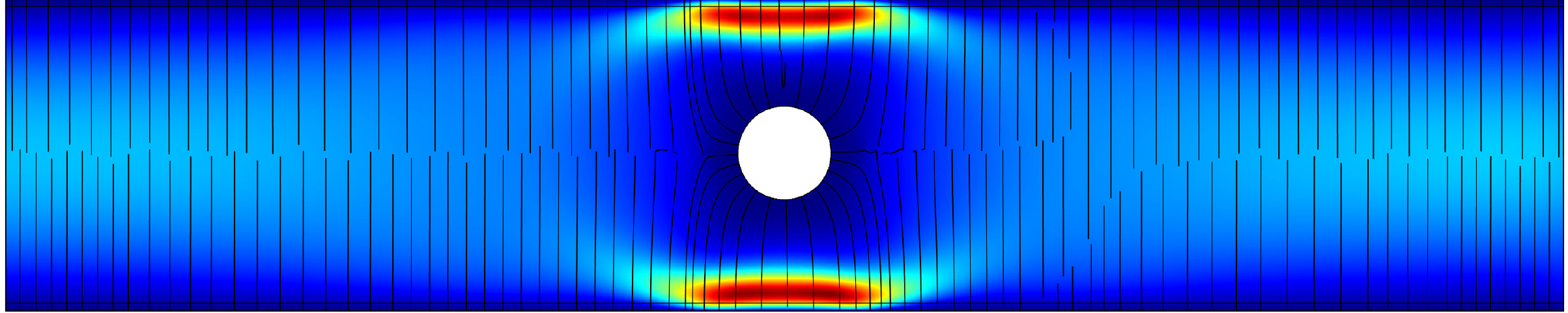}

\vspace{1mm}

\includegraphics[width=0.45\textwidth]{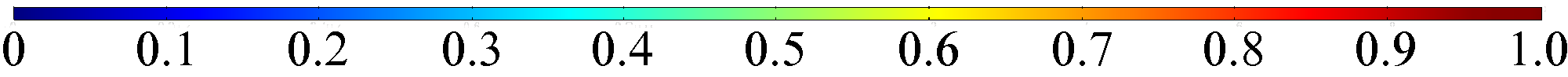}
\caption[.]{Snapshots showing the time evolution of a particle placed inside a NLC microchannel with initial position $Y=0.5$ as it relaxes towards the centre: (a) $T = 0$; (b) $T= 12.5$; and (c) $T= 100$. Here $R=0.3$ and $Er=0.02$. The anchoring conditions on the particle and channel walls are homeotropic. The values for all other parameters used for the calculation are given in Table~\ref{table.params}. }
\label{fig.single.part.motion.snap}
\end{figure}

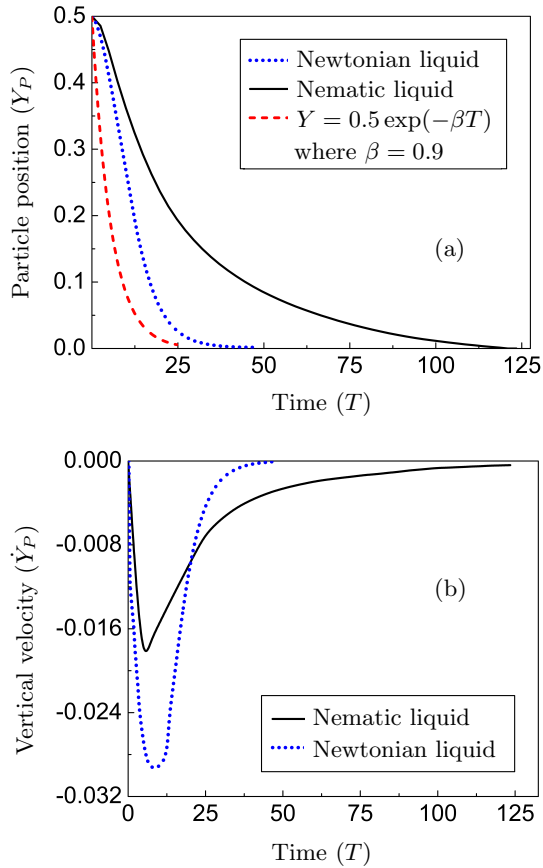
\begin{figure}[!ht]
\begin{overpic}[width=0.4\textwidth]{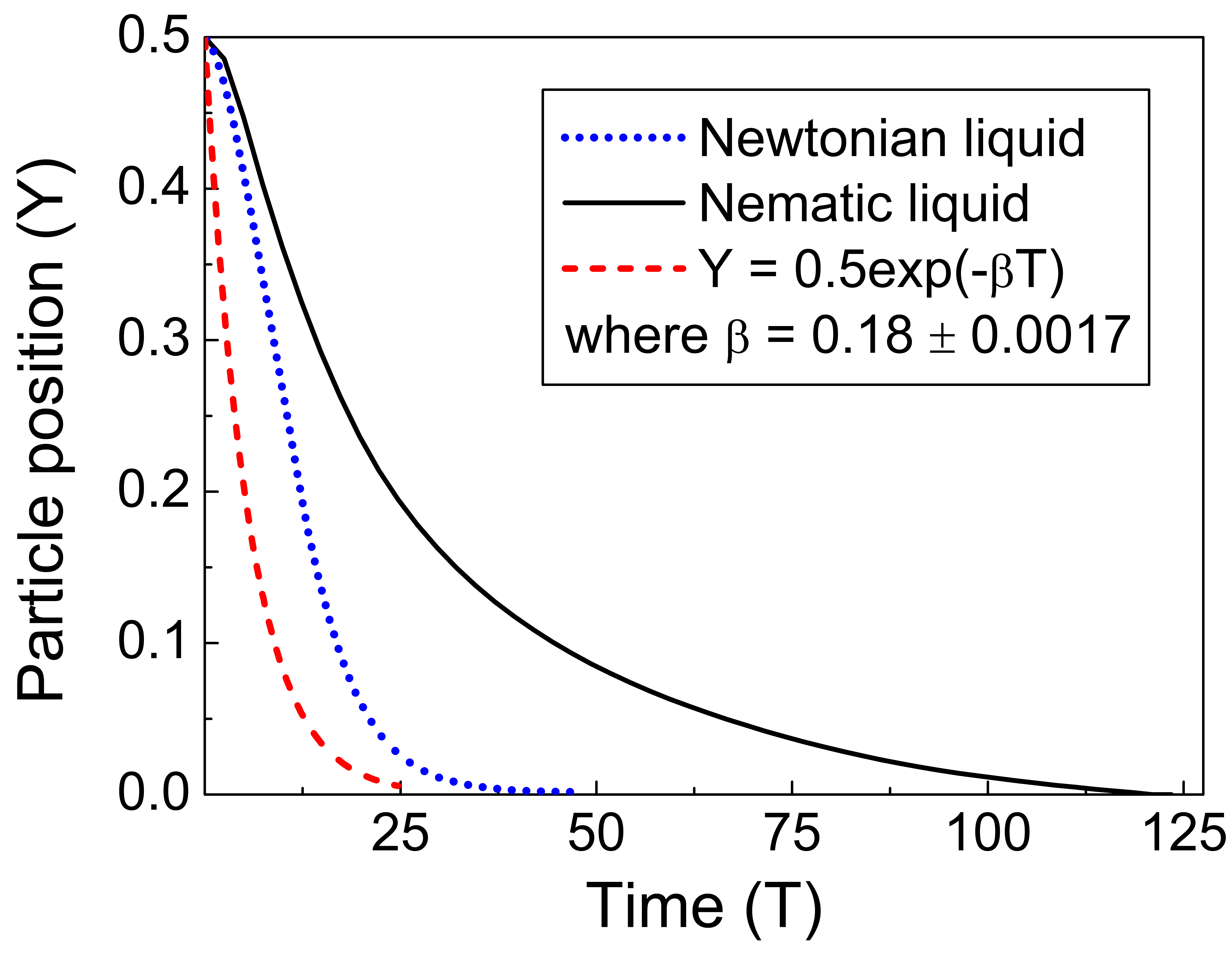}
\put(35,1.5){\colorbox{white}{\makebox[10em]{Time ($T$)}}}
\put(0,15){\rotatebox{90}{\colorbox{white}{\makebox[12em]{Particle position ($Y_P$)}}}}
\put(80,30){(a)}
\put(53,65){\colorbox{white}{Newtonian liquid}}
\put(53,59.5){\colorbox{white}{Nematic liquid}}
\put(53,54){\colorbox{white}{$Y = 0.5\exp (-\beta T)$}}
\put(45,48){\colorbox{white}{~~~~~where $\beta = 0.9$~~~~~~}}
\put(44.5,46.15){\line(1,0){48}}
\end{overpic}

\vspace{4mm}

\begin{overpic}[width=0.4\textwidth]{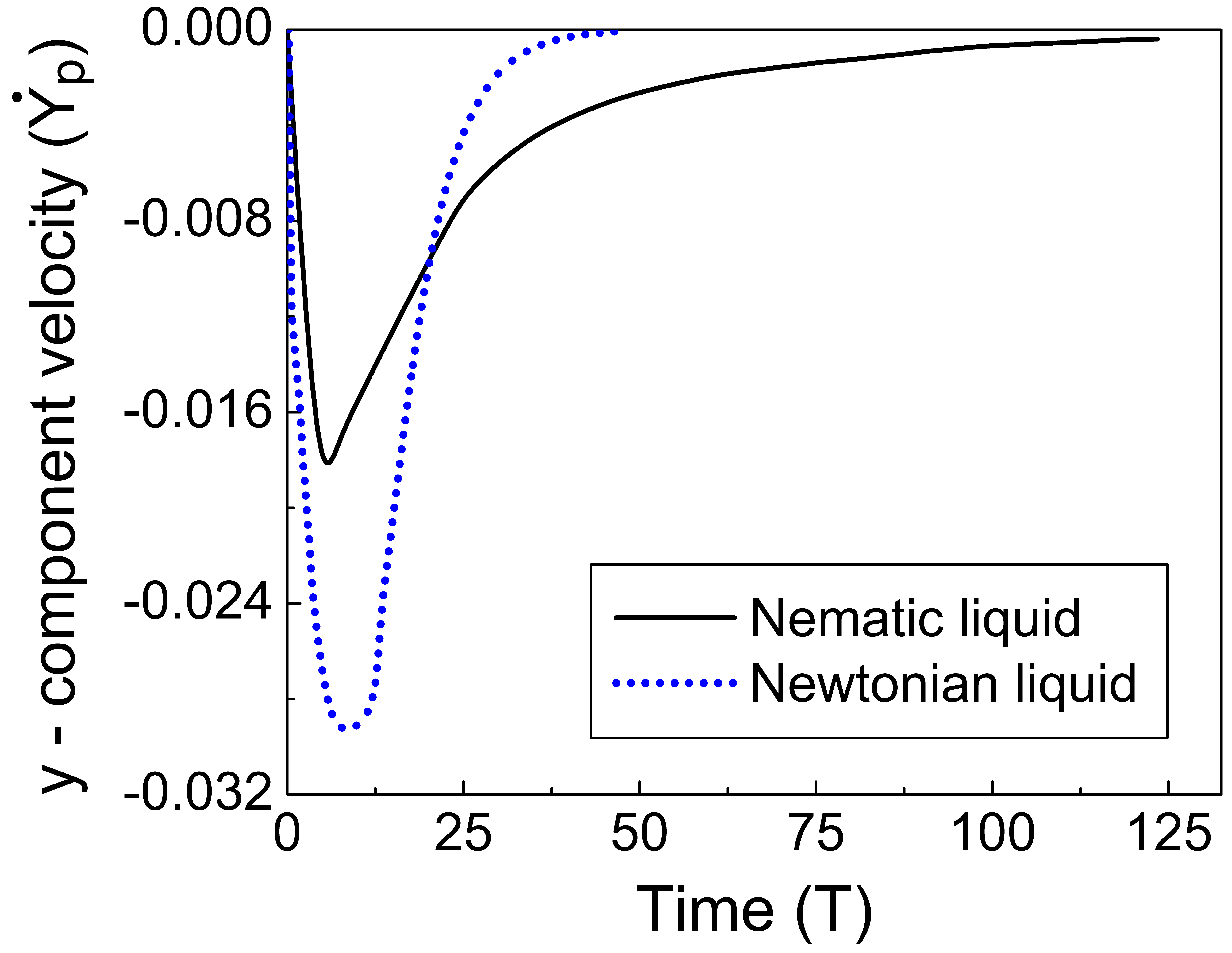}
\put(35,1){\colorbox{white}{\makebox[10em]{Time ($T$)}}}
\put(0,-5){\rotatebox{90}{\colorbox{white}{\makebox[20em]{Vertical velocity ($\dot{Y}_P$)}}}}
\put(80,50){(b)}
\put(56,26){\colorbox{white}{Nematic liquid}}
\put(56,20){\colorbox{white}{Newtonian liquid}}
\end{overpic}

\vspace{0mm}

\caption[.]{Time evolution of a particle (black lines) that begins at position $Y=0.5$ as it relaxes towards the centre $Y=0$ in a NLC: (a) vertical coordinate of particle centre and (b) $Y$-component of the particle velocity. Here $R=0.3$ and $Er=0.02$. The anchoring conditions on the particle and channel walls are homeotropic. The dotted line (in blue) is the behaviour of the particle in a viscous Newtonian flow, obtained by setting the elastic constant, $\kappa = 0$ leading to $Er \rightarrow \infty$. The values taken for all other parameters used for the calculation are given in Table~\ref{table.params}.}
\label{fig.single.part.traj}
\end{figure}

In the Newtonian case the cross-plane position $Y=0$ corresponds to a stable equilibrium: if we release a particle from $Y\neq 0$ then the particle will evolve towards the centre as a result of the viscous stress exerted on the particle~\cite{yang2005migration}. This behaviour is also observed for the values of the NLC parameters considered in Fig. \ref{fig.single.part.motion.snap}, with relaxation approximately following an exponential decay to $Y=0$ (Fig. \ref{fig.single.part.traj}a).  However, there are now two opposing forces that act on the particle: the first, due to the viscous forces, arises as a result of the difference in shear stresses between the bottom and top of the particle (with the lower shear stress on the side closest to the wall), which acts to restore the particle towards the centre (this is the only force in a Newtonian flow) \cite{yang2005migration}; the second arises due to the NLC, which generates an attractive force between the strongly anchored channel wall and the particle surface \cite{tasinkevych2010colloidal}. This attractive force opposes the particle motion towards the centre, increasing the time taken to reach the centre, compared with the case of a Newtonian flow field (Fig. \ref{fig.single.part.traj}a).

Examining the velocity profiles in Fig. \ref{fig.single.part.traj}b, the particle decelerates as it migrates towards the centre. In the case of nematic fluid, the deceleration is slower compared to the Newtonian fluid because of the presence of the elastic forces and interactions with the strongly anchored neighbouring channel wall. 

\begin{figure}
\begin{overpic}[width=0.4\textwidth]{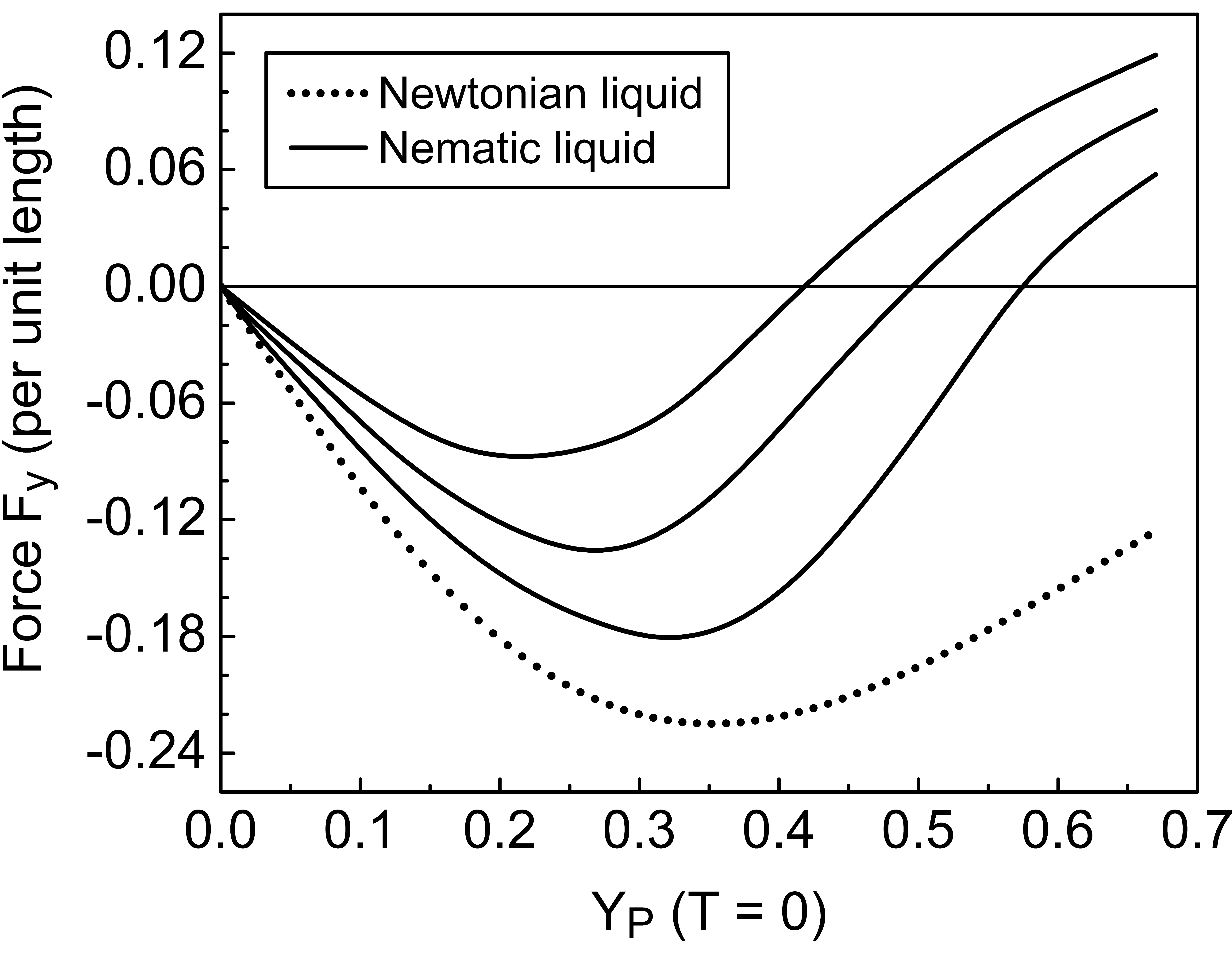}
\thicklines
\centering
\put(35,1){\colorbox{white}{\makebox[10em]{$Y_P ~(T=0)$}}}
\put(-1,10){\rotatebox{90}{\colorbox{white}{\makebox[15em]{Force $F_Y$ (per unit length)}}}}
\put(25,20){(a)}
\put(54,48){\vector(2,-1){25}}
\put(28,49.5){Increasing $Er$}
\end{overpic}

\vspace{5mm}

\begin{overpic}[width=0.4\textwidth]{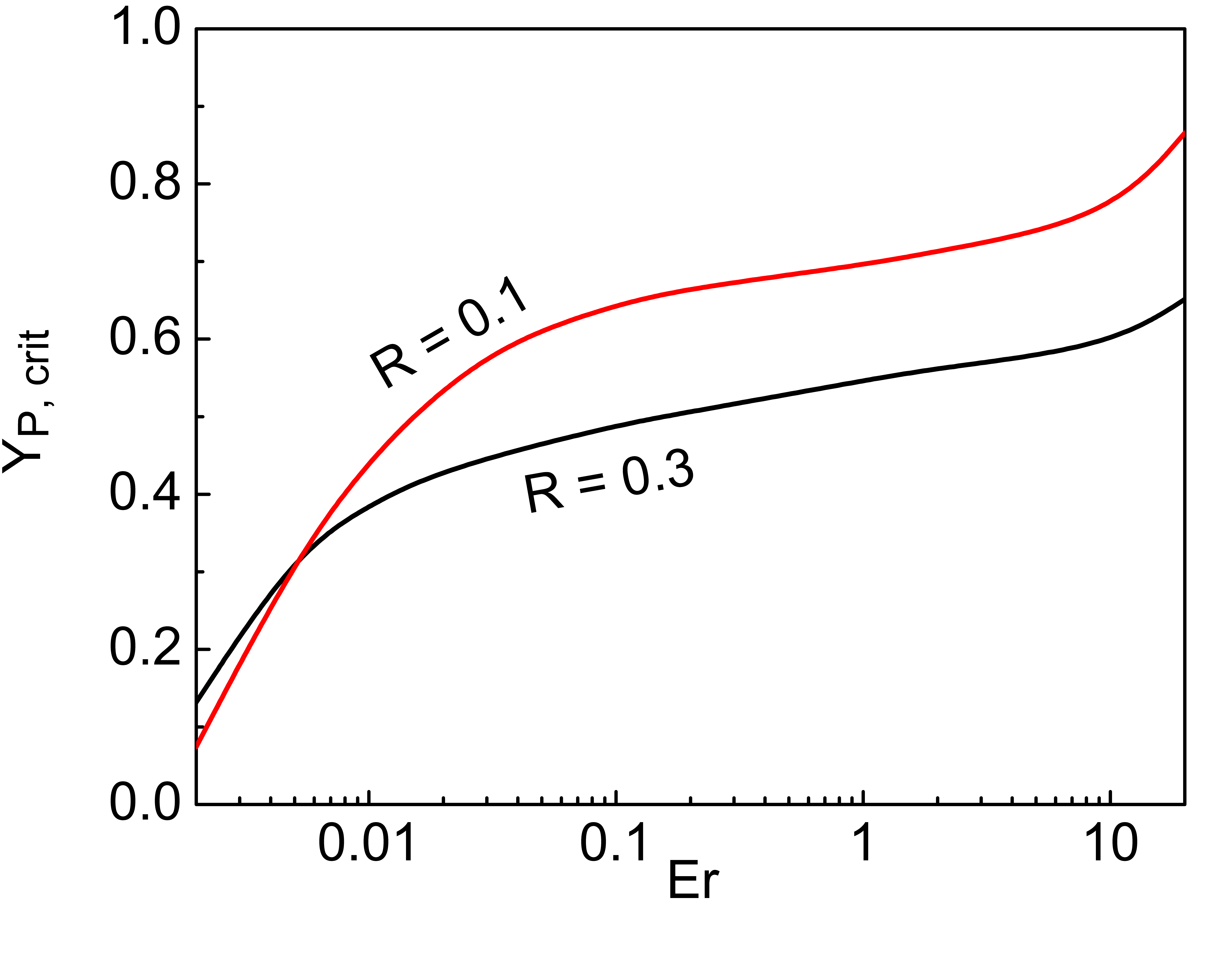}
\centering
\put(45,3.05){\colorbox{white}{\makebox[5em]{$Er$}}}
\put(-0.25,20){\rotatebox{90}{\colorbox{white}{\makebox[10em]{$Y_{P}^\ast$}}}}
\put(25,62){(b)}
\put(27,44){\rotatebox{32}{\colorbox{white}{$R = 0.1$}}}
\put(40,35){\rotatebox{12}{\colorbox{white}{$R = 0.3$}}}
\end{overpic}

\caption{(a) The lift force ($F_Y$) experienced by the particle at different initial $Y$ locations (of the particle centre) in the top half of the channel ($Y > 0$), for $R=0.3$. The three solid curves are for nematic liquid with $Er=$ 0.02, 0.2 and 2. The anchoring conditions on the particle and channel walls are homeotropic. If $F_Y < 0$ the particle migrates towards the centre. For the Newtonian liquid (obtained by setting the elastic constant $\kappa=0$, thus $Er \rightarrow \infty$), the particle migrates towards the centre for all particle radii, $R$~\cite{yang2005migration}. (b) Variation of the critical $Y$ position of the particle, $Y_P^\ast$ (at $T=0$ for which $F_Y = 0$) as a function of the Ericksen number for two different particle sizes. The values taken for all other parameters used for the calculation are given in Table~ \ref{table.params}. The boundary conditions on the particle surfaces and channel walls are homeotropic.}
\label{fig.single.yp.critical}
\end{figure}
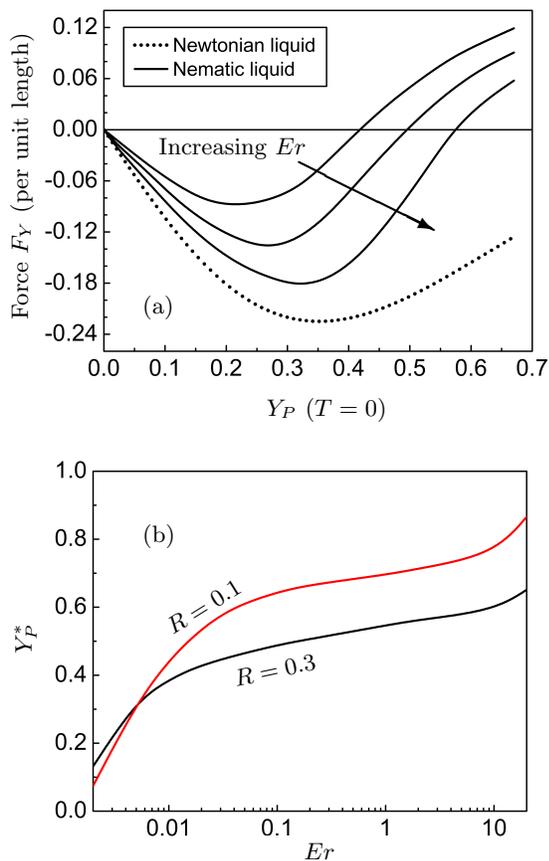

Since there exists a competition between the elastic stresses (due to the NLC) and the viscous forces on the colloidal particle, we hypothesize that there may be a critical vertical position for which the particle might migrate towards the wall instead of the centre. To explore this, we  analyse the lift force on the colloidal particle at $T=0$ to determine whether it migrates towards the centre ($F_Y < 0$ since the particle is located at $Y>0$ initially) or the wall ($F_Y > 0$). An unstable equilibrium is indeed found, at a critical vertical position, $Y_P^\ast$, for which the total lift force is zero (Fig. \ref{fig.single.yp.critical}a). With increasing $Er$, the lift force approaches the limit of the Newtonian flow, where the particle always migrates towards the centre irrespective of its position. The present observation suggests that, with a uniform particle distribution in the channel, the particles located on either side of the critical $Y_P$ would separate out moving either towards the wall or towards the centre. This is somewhat similar to the situation of a Newtonian flow through a channel with porous walls, where the particles tend to flow towards the wall due to the transverse velocity induced by the suction (difference in pressure across the porous wall). However, in this analogy all particles would eventually deposit on (or penetrate) the wall, while in the NLC case only a fraction of the introduced particles would attach to the wall while the remainder would move to the channel centre. The critical $Y$ location is dependent on the size of the particle and Ericksen number (Fig. \ref{fig.single.yp.critical}b). With increasing particle size, the director interaction is stronger therefore reducing the overall lift force, yielding smaller values of $Y_P^\ast$, as observed from Fig. \ref{fig.single.yp.critical}b.

\subsubsection{Dual particle system}

\begin{figure}[!ht]
\includegraphics[width=0.45\textwidth]{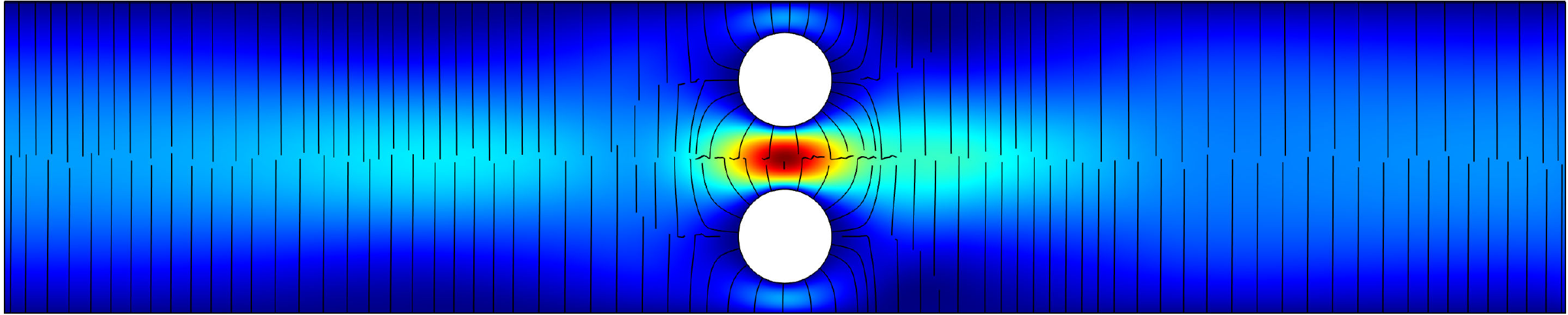}

\vspace{1mm}

\includegraphics[width=0.45\textwidth]{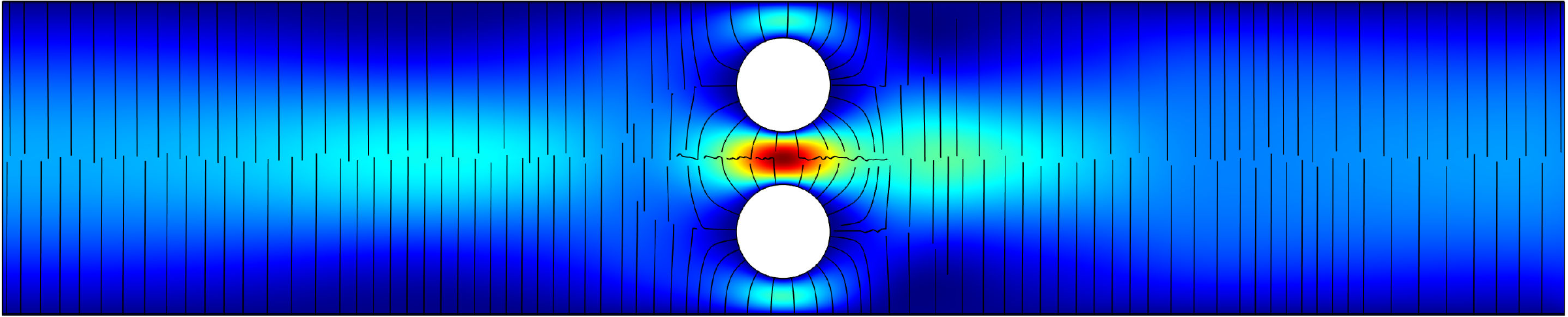}

\vspace{1mm}

\includegraphics[width=0.45\textwidth]{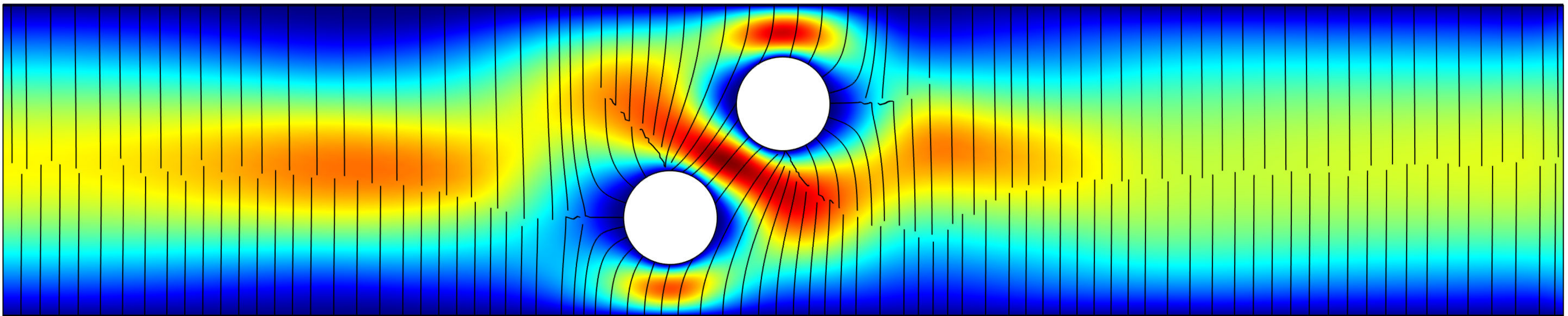}

\vspace{1mm}

\includegraphics[width=0.45\textwidth]{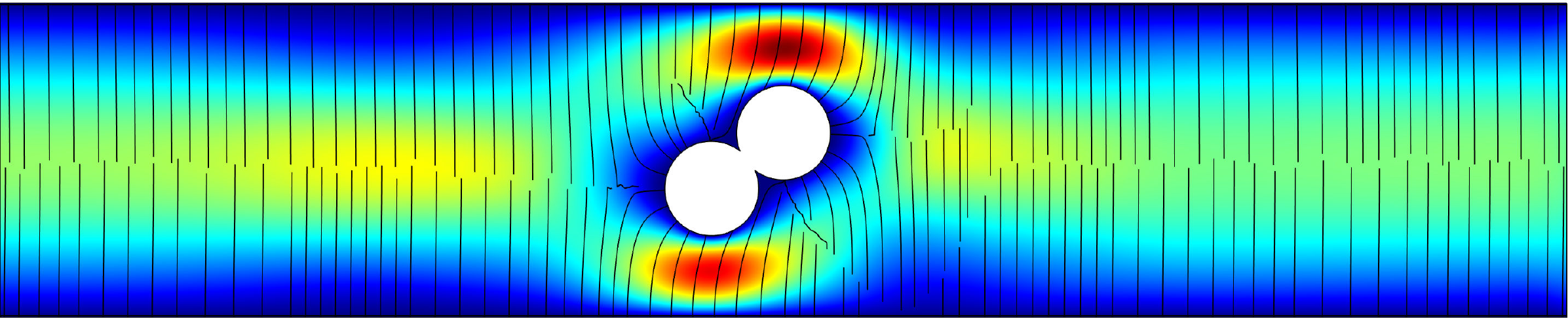}

\vspace{1mm}

\includegraphics[width=0.45\textwidth]{traj_color_bar.png}
\caption[.]{Snapshots showing the evolution with time of two particles placed in a microchannel located at $X = 0, Y=\pm 0.5$, as they approach each another and agglomerate: (a) $T = 0$; (b) $T= 0.125$; (c) $T=0.5$; and (d) $T=1.5$ (at which point the particles have reached their equilibrium configuration). Here $R=0.3$ and $Er=0.02$. The anchoring conditions on the particle and channel walls are homeotropic. The values taken for all other parameters used for the calculation are given in Table~\ref{table.params}. The boundary conditions on the particle surfaces and channel walls are homeotropic.}
\label{fig.double.part.motion.snap}
\end{figure}

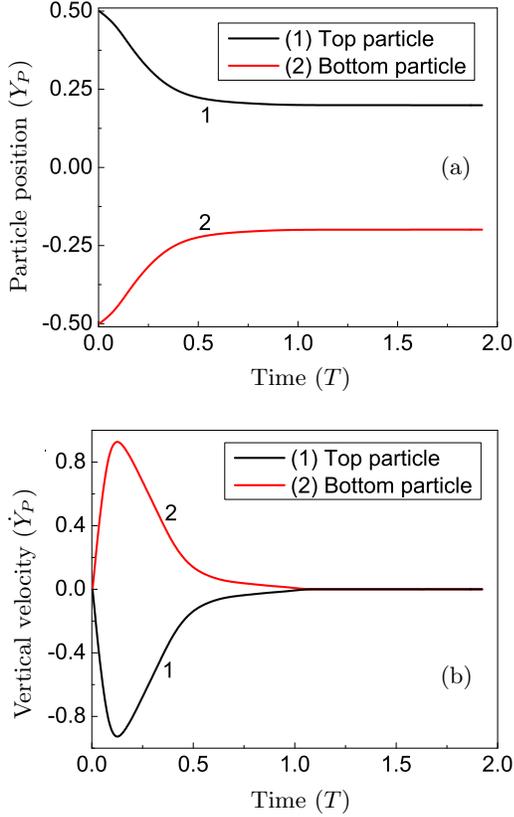
\begin{figure}[!ht]
\begin{overpic}[width=0.4\textwidth, trim={0 4cm 4cm 0},clip]{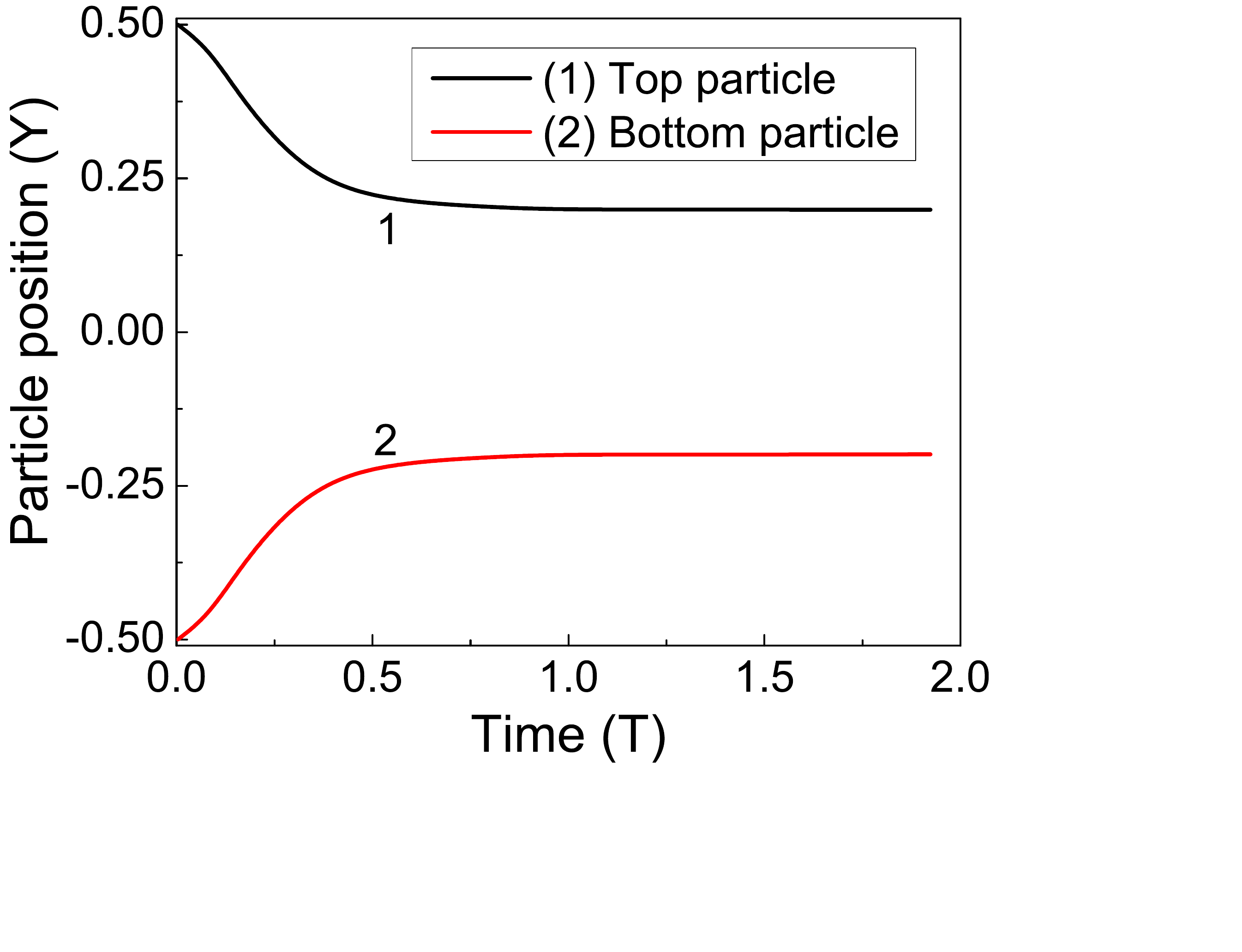}
\put(30,1){\colorbox{white}{\makebox[10em]{Time ($T$)}}}
\put(-1.5,-7){\rotatebox{90}{\colorbox{white}{\makebox[20em]{Particle position ($Y_P$)}}}}
\put(80,40){(a)}
\end{overpic}

\vspace{5mm}

\begin{overpic}[width=0.4\textwidth, trim={0 4cm 4cm 0},clip]{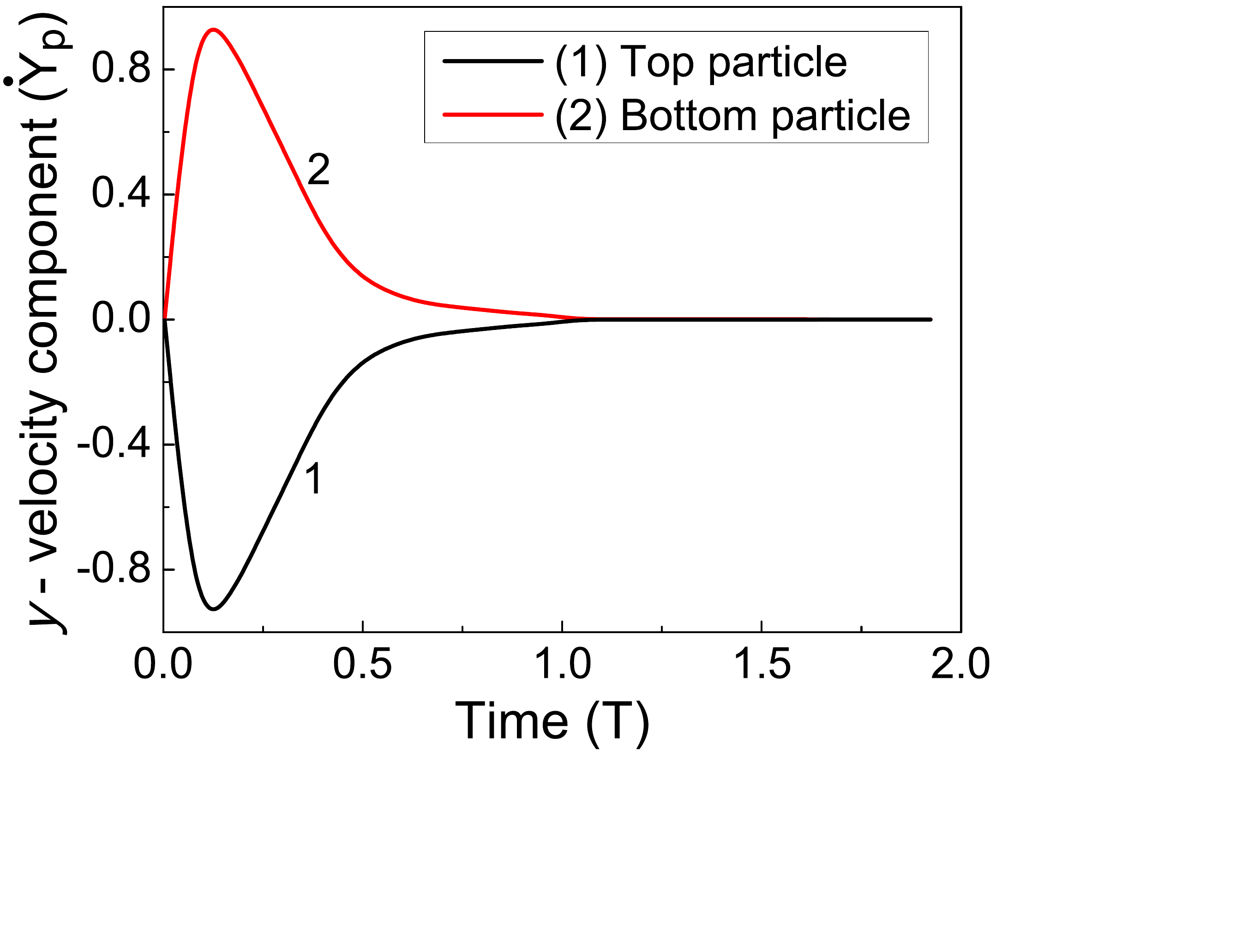}
\put(30,2){\colorbox{white}{\makebox[10em]{Time ($T$)}}}
\put(-1.5,-7){\rotatebox{90}{\colorbox{white}{\makebox[20em]{Vertical velocity ($\dot{Y}_P$)}}}}
\put(80,25){(b)}
\end{overpic}

\vspace{1mm}

\caption[.]{Time evolution of two particles that start at positions $Y=\pm 0.5$, relaxing towards the equilibrium configuration: (a) vertical coordinates of the particle centres; and (b) $Y$-component of the particle velocities. Here $R=0.3$ and $Er=0.02$. The anchoring conditions on the particle and channel walls are strongly homeotropic. The values taken for all other parameters used for the calculation are given in Table~\ref{table.params}. The boundary conditions on the particle surfaces and channel walls are homeotropic.}
\label{fig.double.part.traj}
\end{figure}

When two particles are released side by side (with zero initial velocities $\dot{X}_P=\dot{Y}_P=0$ starting from $X=0, Y=\pm 0.5$) within the microchannel, the presence of the attractive NLC forces between two homeotropically anchored particle surfaces results in significantly increased velocities in the $Y$ direction, with the particles eventually touching one another (Fig. \ref{fig.double.part.motion.snap}, \ref{fig.double.part.traj}). Upon agglomeration, the overall drag force is adjusted according to Fig. \ref{fig.forces}c. We observe that there are two stages in the dynamics of two particles inside a NLC microchannel. The first stage corresponds to the two isolated particles attracting each other and moving towards each other in a straight line ($T \lesssim 0.35$). In the second stage, the agglomerate reorients due to the quadrupolar interactions ($ 0.35 \lesssim T \lesssim 1.1$) (Fig. \ref{fig.double.part.traj}) \cite{turiv2013effect, Lavrentovich}. Due to the presence of quadrupolar interactions \cite{turiv2013effect, Lavrentovich}, the particles tend to reorient themselves with the angle of inclination $\theta_p\approx 39^o$, with respect to the horizontal axis $Y=0$ measured in the anti-clockwise direction. This is corroborated by Mondiot et al. \cite{mondiot2014colloidal} who find an angle of inclination $\theta_p = \sqrt{\arccos(4/7)} \approx 40.9^o$, obtained by minimizing the quadrupolar interaction energy. This colloidal particle re-orientation is driven by the minimum energy configuration state as described in \cite{senyuk2016hexadecapolar}.
The experimental observations presented in the literature also support the attractive force in the direction of the quadrupolar interactions \cite{tasinkevych2010colloidal, al2004forces, takahashi2008direct, takahashi2008force}. The calculation of the Landau-de Gennes free energy of the two isolated particles suggests that the minimum energy depends on $\theta_p$ and the separation distance \cite{tasinkevych2010colloidal}. Essentially, one would expect a re-orientation and re-alignment of the dual particle system as the separation gap closes, as is observed in our numerical experiment.

\subsubsection{Triple-particle system}

\begin{figure}[!ht] 
\includegraphics[width=0.45\textwidth]{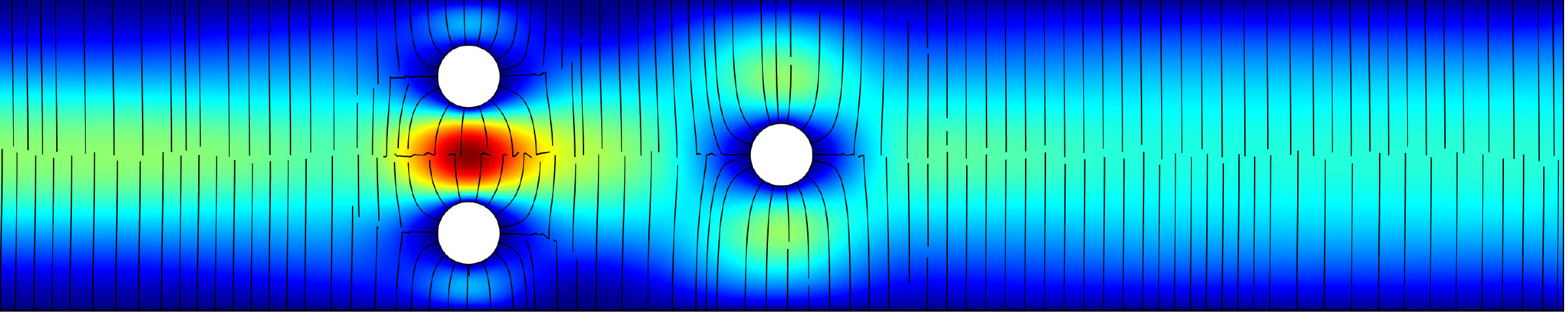}

\vspace{1mm}

\includegraphics[width=0.45\textwidth]{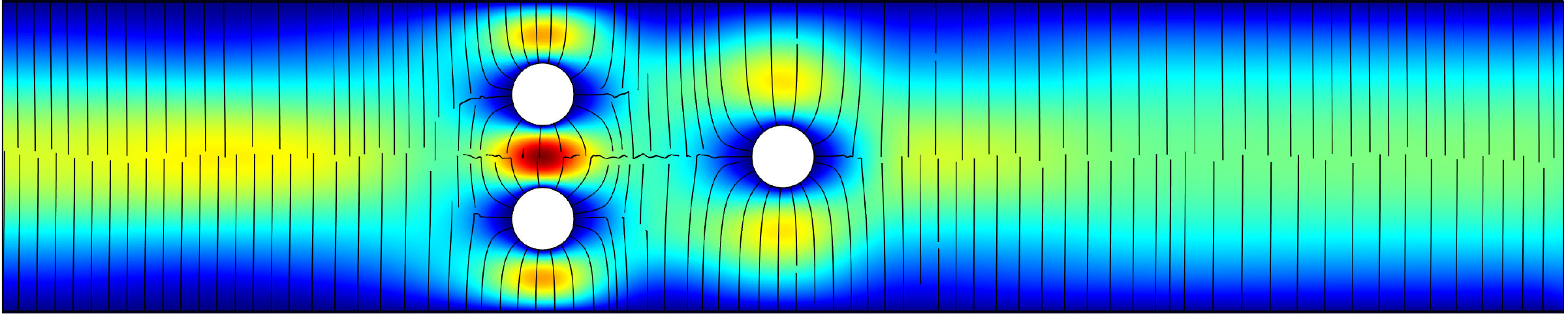}

\vspace{1mm}

\includegraphics[width=0.45\textwidth]{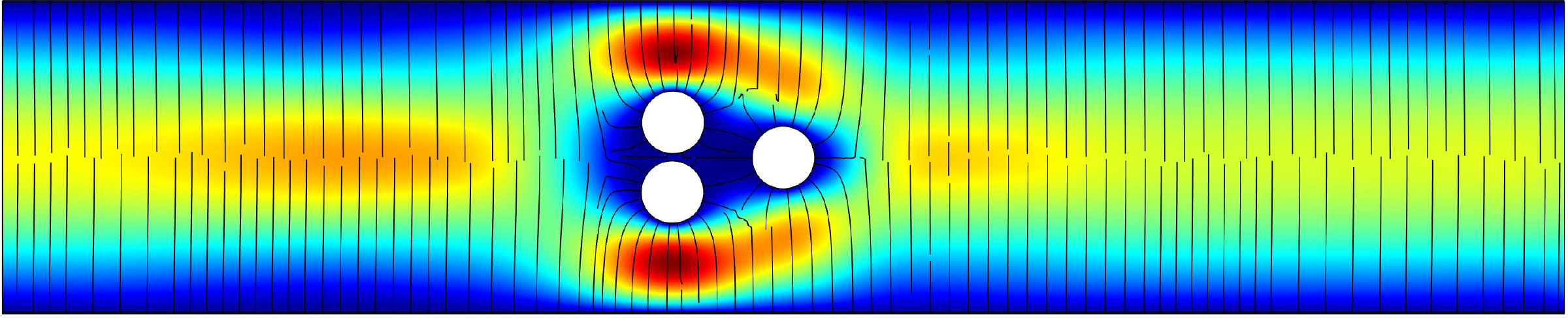}

\vspace{1mm}

\includegraphics[width=0.45\textwidth]{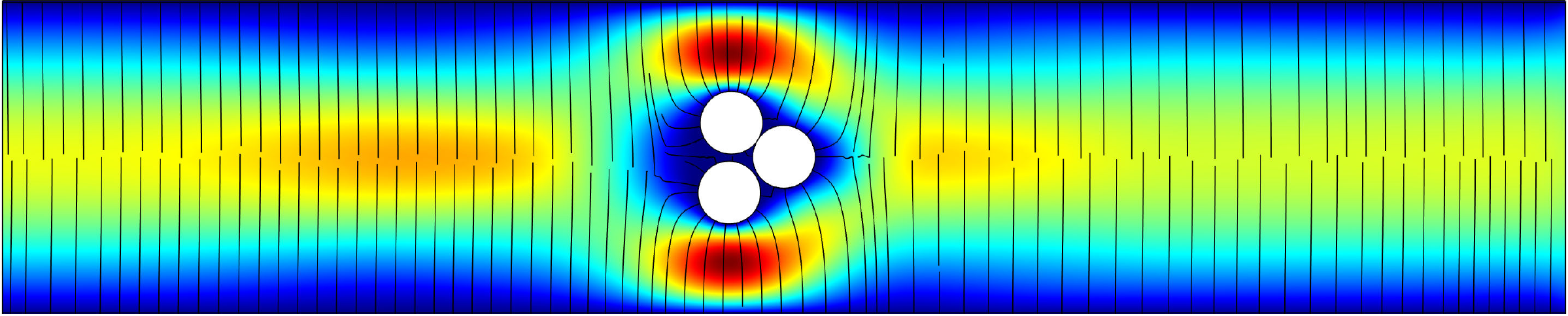}

\vspace{1mm}

\includegraphics[width=0.45\textwidth]{traj_color_bar.png}
\caption[.]{Snapshots showing the evolution with time of three particles placed in a microchannel located at $X=-0.2$, $Y=\pm 0.5$ and $X=0$, $Y=0$: (a) $T = 0$; (b) $T= 0.1$; (c) $T=0.3$; and (d) $T=0.5$ (at which point the three-particle system has reached equilibrium). Here $R=0.2$ and $Er=0.02$. The anchoring conditions on the particle and channel walls are strongly homeotropic. All other parameters used for the calculation are given in Table~\ref{table.params}. The boundary conditions on the particle surfaces and channel walls are homeotropic.}
\label{fig.triple.part.motion.snap}
\end{figure}

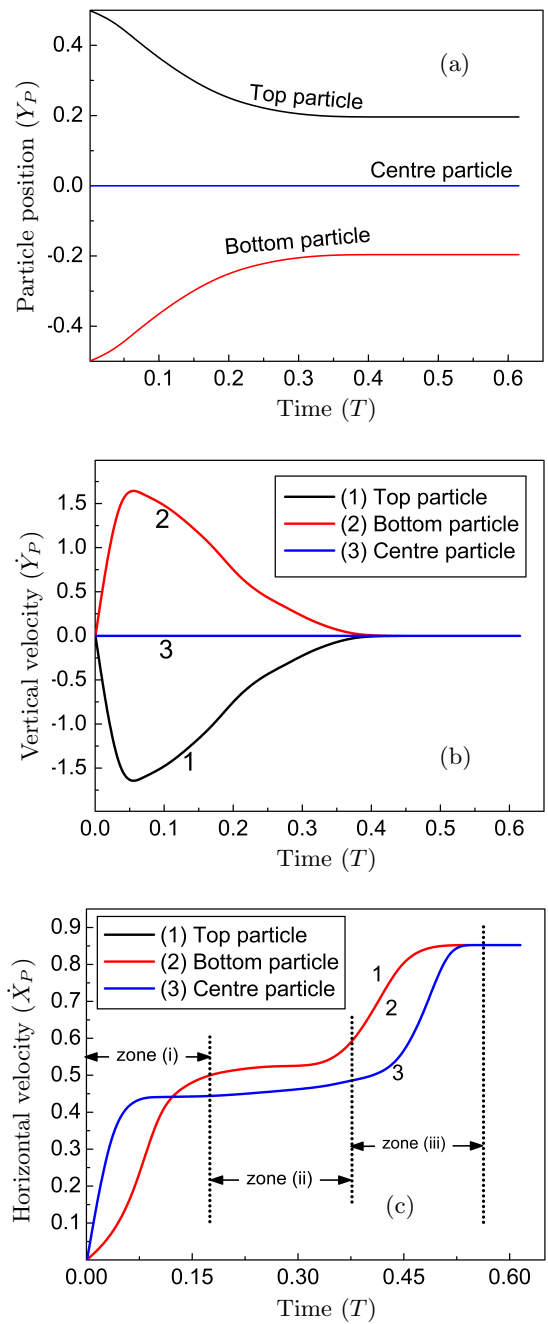
\begin{figure}[!ht]
\begin{overpic}[width=0.4\textwidth]{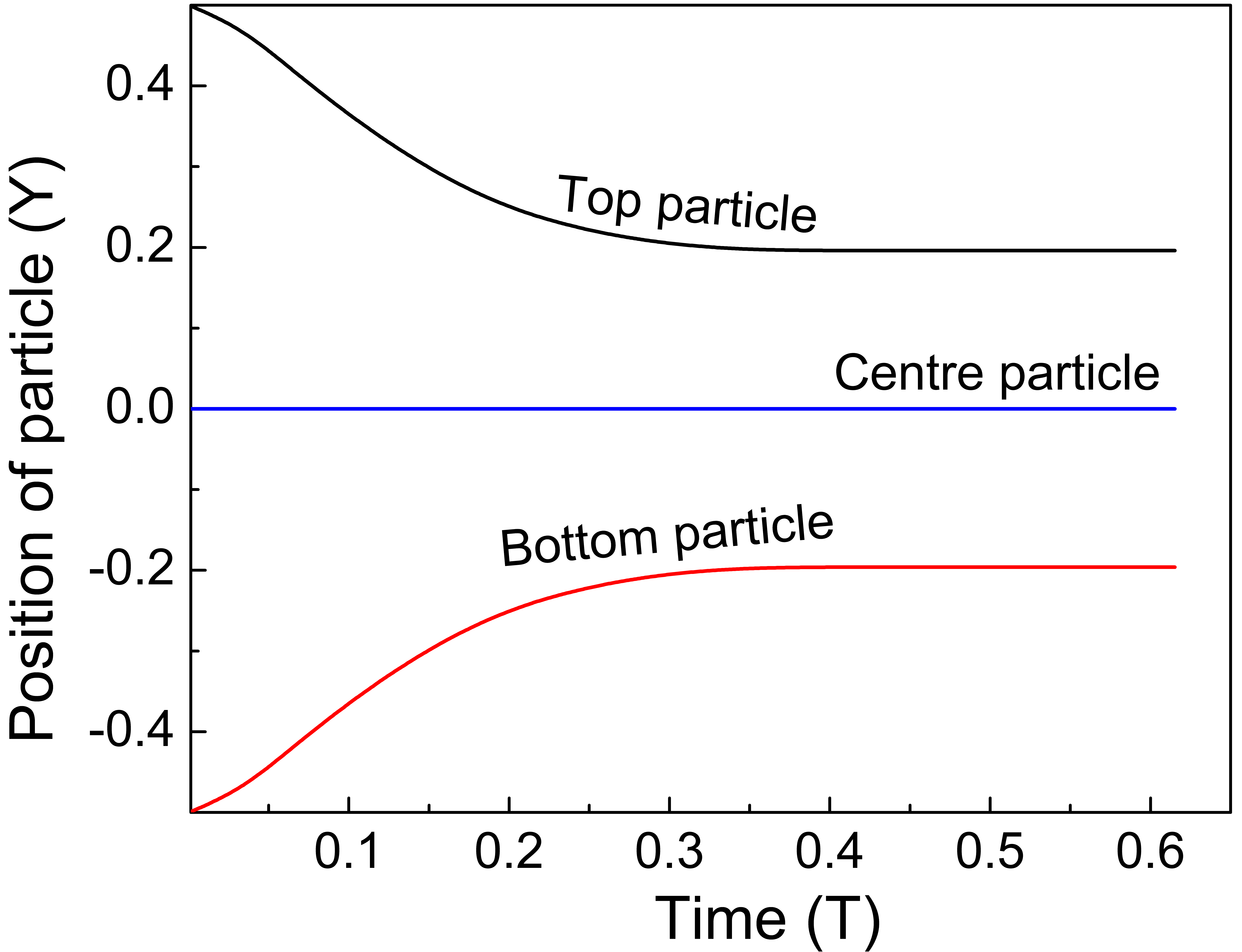}
\put(35,1){\colorbox{white}{\makebox[10em]{Time ($T$)}}}
\put(0,-5){\rotatebox{90}{\colorbox{white}{\makebox[20em]{Particle position ($Y_P$)}}}}
\put(80,65){(a)}
\end{overpic}

\vspace{5mm}

\begin{overpic}[width=0.4\textwidth]{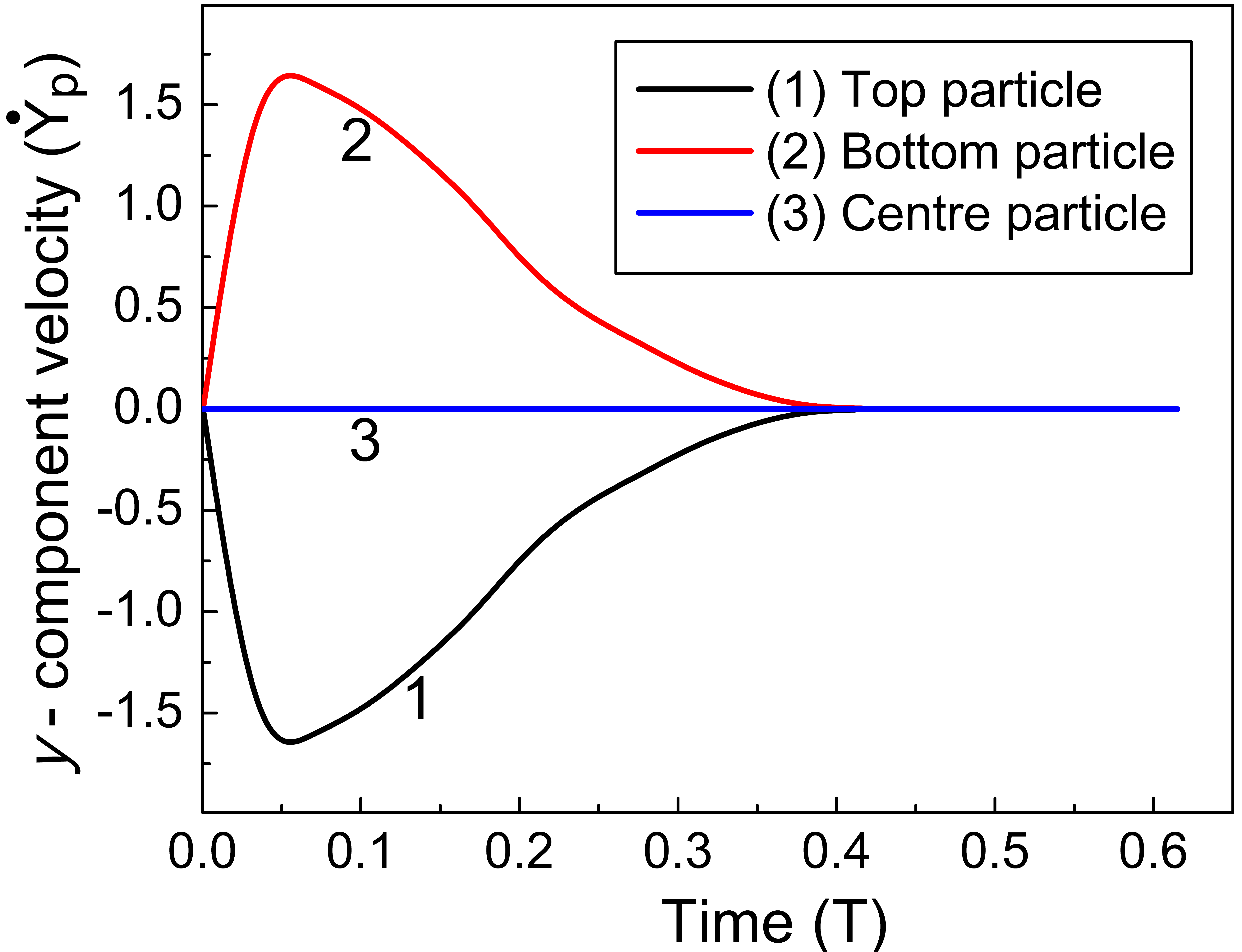}
\put(35,1){\colorbox{white}{\makebox[10em]{Time ($T$)}}}
\put(0,-5){\rotatebox{90}{\colorbox{white}{\makebox[20em]{Vertical velocity ($\dot{Y}_P$)}}}}
\put(80,20){(b)}
\end{overpic}

\vspace{5mm}

\begin{overpic}[width=0.4\textwidth]{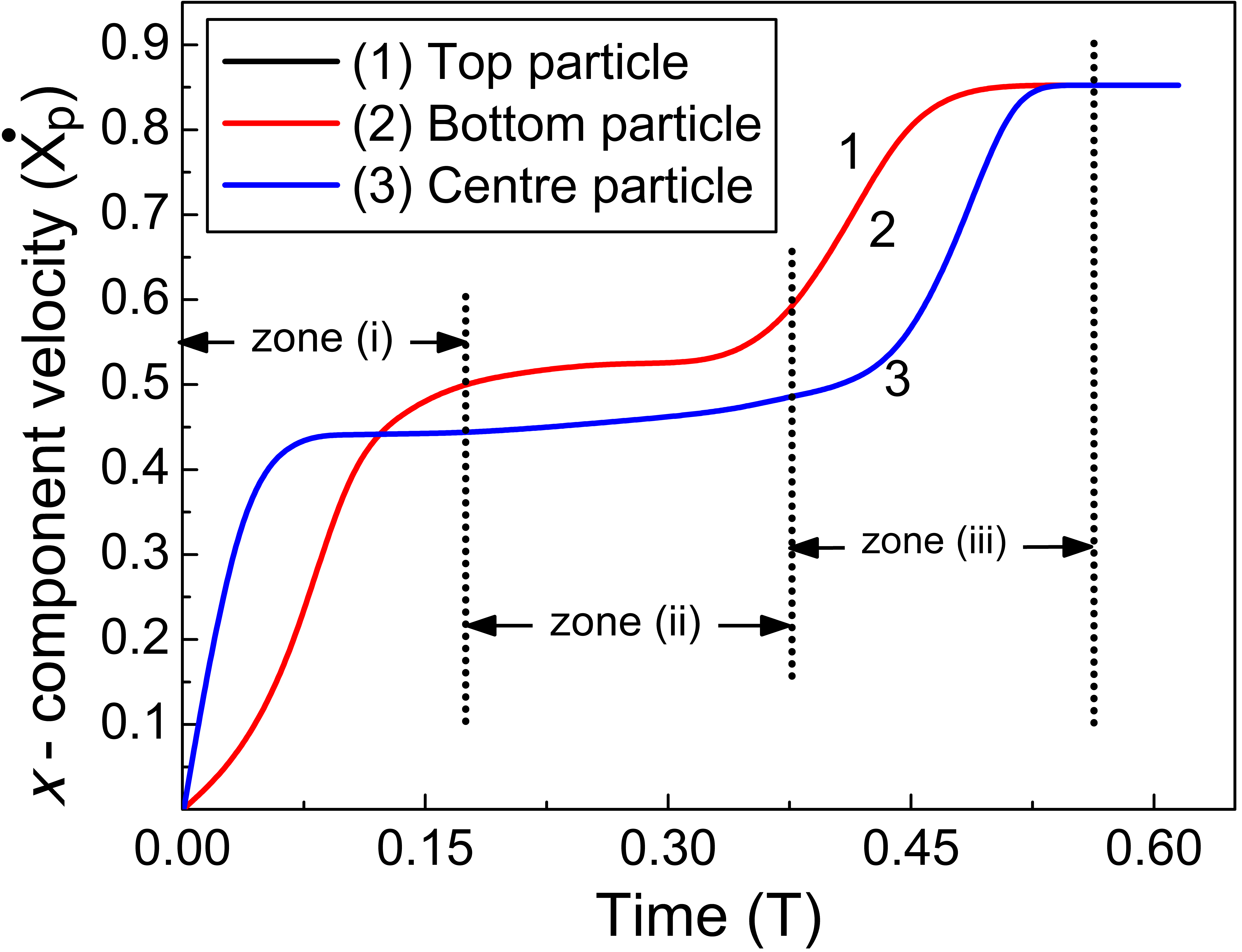}
\put(35,1){\colorbox{white}{\makebox[10em]{Time ($T$)}}}
\put(-1.5,-4){\rotatebox{90}{\colorbox{white}{\makebox[20em]{Horizontal velocity ($\dot{X}_P$)}}}}
\put(70,20){(c)}
\end{overpic}

\vspace{0mm}

\caption[.]{Time evolution of three particles that begin at positions $X=-2$, $Y=\pm 0.5$ and $X=0$, $Y=0$, relaxing towards the equilibrium: (a) vertical coordinates of particle centres; (b) $Y$-components of the particle velocities and (c) $X$-components of the particle velocities. Here $R=0.2$ and $Er=0.02$. The anchoring conditions on the particle and channel walls are strongly homeotropic. All other parameter values used for the calculation are given in Table~\ref{table.params}.}
\label{fig.triple.part.traj}
\end{figure}

Finally, we consider the nematohydrodynamic effects on the mechanics of a triple-particle system (Fig. \ref{fig.triple.part.motion.snap}), each having radii $R=0.2$. We begin with a configuration in which one particle is placed at the centre $X=Y=0$ while the other two particles are placed at positions $X=-2,Y=\pm0.5$. Initially all the particles are held stationary ($\dot{X}_P = \dot{Y}_P = 0$ at $T = 0$). For early times (for $T \lesssim 0.1$) the two particles initially at $X=-2, Y=\pm0.5$, aggregate before approaching the third (central) particle (Fig. \ref{fig.triple.part.motion.snap}a,b,c). As the two-particle agglomerate approaches (at a higher velocity compared to the central particle), the central particle accelerates before the agglomerate binds to the central particle (Fig. \ref{fig.triple.part.traj}c). When the two-particle agglomerate finally catches up with the single particle, the particles align themselves as an equiangular system (joining the centres forms an equilateral triangle) to minimize the overall energy, due to the quadrupolar interactions (Fig. \ref{fig.triple.part.motion.snap}d). 

There are three distinct velocity zones observed in this figure.

(i) For $T \lesssim 0.16$, the two off-centred particles approach one 
another while accelerating towards the third, central, particle. After 
these two particles aggregate the overall drag driving the particle in the $X$ direction increases (as 
predicted by Fig. \ref{fig.forces}c), since the cumulative size of the dual-particle agglomerate is less than the critical size in Fig.\ref{fig.forces}c.  The isolated central third particle is largely unaffected by the two other particles during this process.

(ii) For $0.16  \lesssim T  \lesssim 0.375$, the agglomerate 
approaches the isolated particle at a higher $X$-velocity than
the central particle. Since the timescale of the final re-orientation of the dual-particle system (due to the quadrupolar interactions) is longer (formed after $T \approx 1$ as seen in Fig. \ref{fig.double.part.traj}a), the two-particle globule catches up with the central particle before the transitional reorientation can occur as observed in the dual-particle system.

(iii) For $T \gtrsim 0.375$, the three particles attach to each other to form a triple agglomerate that moves as a whole. We note that a triplet formed of individual particles (of size $R$) can be treated to have an effective size of $(1+\sqrt{3})R \approx 0.54$ (for $R = 0.2$), which is below the size that maximizes the drag, $R_{max} = 0.56$ in Fig. \ref{fig.forces}c. This results in an increased velocity of the agglomerate due to the enhanced drag as seen by Fig. \ref{fig.forces}c, by considering the effective size of the aggregate relative to individual particle sizes.

\clearpage
\newpage

\vspace{4cm}

\section{Conclusions}
We have conducted a batch of numerical experiments on particles immersed in a NLC microfluidic cell, motivated by recent experimental work, to investigate the role of the anchoring on particle boundaries, the particle size and the Ericksen number in both the transient dynamics and the long-term dynamics. In the single-particle case, we numerically studied the lift force on particle inclusions and how this force may be used for particle separation, with particle deposition on the walls and particle movement towards the channel centre as a function of initial locations. While the reorientation of a two-particle globule in a NLC environment has been reported in the literature, we investigate the dynamics: the attraction stage, the attachment stage, and the reorientation stage. One could perform parametric sweeps with different initial conditions or different initial locations of the two or three particles to determine the solution landscape. It is not \emph{a priori} clear how the particles will approach each other, for example if the two particles do not begin vertically on top of each other as in our example, they might approach the centre and then assemble as a linear chain propagating along the channel. The three-particle dynamics is much richer. If the two-particle globule has time to reorient before attaching to the third central particle (the two-particle globule does not have time to reorient in our example), then the final three-particle globule may have a different shape or different dynamics. The examples presented here have illustrated a variety of phenomena that are not observed in a Newtonian fluid. The results demonstrate the rich hydrodynamic landscape for NLC microfluidics and how the coupling between flow, anchoring, particle sizes and NLC order may tune or control experimentally observed phenomena.

\section*{Acknowledgements}
The authors gratefully acknowledge helpful discussions with L.~J.~Cummings and D.~Vigolo. AM and IMG are grateful for discussions on nematic microfluidics with A.~Sengupta and for the UK Fluids Network Short Research Visit scheme for facilitating our collaboration. SM is grateful to EPSRC and the Royal Society for financial support. AM's research is supported by an EPSRC Career Acceleration Fellowship EP/J001686/1 and EP/J001686/2, an OCIAM Visiting Fellowship and the Advanced Studies Centre at Keble College. IMG gratefully acknowledges support from the Royal Society through a University Research Fellowship.

\newpage

\bibliographystyle{unsrt}
\bibliography{sourav_bibtex_refs}

\end{document}